\documentclass[
  aps,
  pre,
  reprint,
  superscriptaddress,
  amsmath,
  amssymb,
  floatfix,
  nofootinbib
]{revtex4-2}
\usepackage{amsmath,amssymb,amsthm}
\usepackage{graphicx}
\usepackage[dvipsnames]{xcolor}
\usepackage[colorlinks=true,allcolors=MidnightBlue]{hyperref}
\usepackage{comment}
\usepackage{braket}
\usepackage{placeins}

\newcommand{\Z}{\mathbb{Z}}
\newcommand{\Sym}{\mathbb{S}}

\newcommand{\cZ}{\mathcal{Z}}
\newcommand{\cE}{\mathcal{E}}
\newcommand{\cS}{\mathcal{S}}
\newcommand{\cM}{\mathcal{M}}

\newcommand{\cB}{\mathcal{B}}
\newcommand{\GFP}{G_{F,P}}
\newcommand{\Aut}{\operatorname{Aut}}

\theoremstyle{plain}
\newtheorem{theorem}{Theorem}
\newtheorem{proposition}{Proposition}

\begin{document}

\title{State-space symmetry in Boolean network ensembles}

\author{Yi-Kai Mo}
\email{ymo6@uh.edu}
\affiliation{Department of Physics, University of Houston, Houston, Texas 77204-5005, USA}
\affiliation{Texas Center for Superconductivity, University of Houston, Houston, Texas 77204-5002, USA}

\author{Kevin E. Bassler}
\email{bassler@uh.edu}
\affiliation{Department of Physics, University of Houston, Houston, Texas 77204-5005, USA}
\affiliation{Texas Center for Superconductivity, University of Houston, Houston, Texas 77204-5002, USA}
\affiliation{Department of Mathematics, University of Houston, Houston, Texas 77204-3008, USA}

\date{\today}

\begin{abstract}
Studies of evolving Boolean network systems have found that Boolean functions cluster into equivalence classes with a characteristic frequency distribution — a pattern governed by the Zyklenzeiger symmetry group — but the dynamical origin of this symmetry has remained unexplained. Here we show that it is a consequence of a fundamental symmetry of Boolean network dynamics itself. A Boolean network consists of nodes that update according to Boolean functions of their regulatory inputs; its global dynamics is encoded in a state-space network. We identify the dynamical symmetry group of ensembles of Boolean networks that average over random interaction topologies: for networks with $N$ nodes, it is the $N$-dimensional hyperoctahedral group, generated by node-flips and node-label permutations. These transformations are admissible for every Boolean network regardless of topology; for random-topology ensembles, including all ensembles of fixed in-degree $K<N$, they constitute the complete symmetry group. We also show that input-label permutations constitute a hidden gauge freedom of the dynamics. The Zyklenzeiger symmetry of Boolean functions follows directly from the node-flips of the hyperoctahedral group together with this gauge freedom. Boolean networks related by the hyperoctahedral group have isomorphic state-space networks and form its orbits, and Burnside’s lemma gives the number of these orbits for any fixed network topology. It provides a symmetry reduction of the analysis of Boolean network dynamics and explains why the Boolean functions of evolved networks fall into Zyklenzeiger classes, and why these classes split in finite networks.
\end{abstract}

\maketitle

\section{Introduction}
\label{sec:intro}
Symmetry is one of the central tools of physics. It allows a simpler and fuller understanding of physical laws and processes. In classical mechanics, continuous symmetries of the action give rise to conservation laws and provide the foundation for geometric and symplectic reduction~\cite{noether_invariant_1918,marsden_ratiu_1999}. In special relativity, Lorentz symmetry unites space and time into a single spacetime~\cite{Minkowski_1909}. In quantum mechanics, symmetry produces conservation laws and selection rules~\cite{tinkham_group_1964}, and in quantum field theory, gauge symmetry determines the form of the fundamental interactions and the particles that carry them~\cite{yang_mills_1954,weinberg_leptons_1967}. In statistical mechanics, the symmetry of the order parameter, together with the spatial dimension, determines the universality class of critical behavior, and the breaking of symmetry distinguishes phases of matter~\cite{landau_phase_1937,wilson_kogut_1974}. Symmetry breaking also organizes biological and chemical dynamics, from reaction--diffusion patterns and cell polarity~\cite{turing1952chemical,Maini1997SpatialPatterns,Goryachev2008Cdc42} to the ordered phases of flocks~\cite{toner_flocks_1998,solon_flocking_2015}, and equivariant bifurcation theory uses symmetry to classify solution branches and the transitions between them~\cite{golubitsky_stewart_schaeffer_1988}. Across these fields, symmetry reduces a problem by identifying equivalent states, and the breaking of symmetry marks qualitative changes in behavior.

Boolean networks (BNs), in which each node is a binary variable updated by a Boolean function (BF) of the states of its inputs, have mathematical roots that predate their use as models of biological dynamics. Boole developed an algebraic approach to logic~\cite{Boole1847}, which was extended by Schröder and others~\cite{Schroeder1890, Couturat1905}. Ehrenfest proposed applying the algebra of logic to the analysis of automatic telephone exchanges~\cite{Ehrenfest1910}, and a theory of switching circuits was later developed independently by Nakashima, Shannon, and Shestakov~\cite{Nakashima1936, shannon_symbolic_1938, Shestakov1941}; see also~\cite{Brown1990}. Turing’s “unorganized machines,” randomly connected networks of two-input NAND units updated in synchrony, anticipated random BNs~\cite{Turing1969}. Harrison gave a systematic treatment of Boolean algebra, switching functions, sequential circuits, and finite automata~\cite{Harrison1965}. Kauffman introduced random BNs as models of gene regulation~\cite{Kauffman1969}, and BNs are now widely used to model signaling, cell-cycle control, and other biological processes~\cite{li_yeast_2004,Garg2008,davidich_boolean_2008,irons_logical_2009,abou_jaoude_logical_2016,kadelka_meta-analysis_2024}.
More recently, Cheng and co-workers developed a matrix formulation of BN dynamics based on the semi-tensor product~\cite{cheng2011analysis}.
Studies of the evolution of BNs have been used to explore the emergence of canalization and criticality in gene regulation \cite{Bornholdt2000,bassler_evolution_2004,liu_emergent_2006,reichhardt_canalization_2007,szejka_evolution_2010,liu_finite_2011,matthew_d_reichl_canalization_2011,MarquesPita2013,shabnam_hossein_symmetry_2014}. BNs are also used to study information processing and control in computer science and engineering \cite{burch_symbolic_1992,cheng2011analysis,snyder_computational_2013}.

In network science, symmetry is studied primarily as a property of the interaction graph. Graph automorphisms partition structurally indistinguishable nodes into classes with identical functional roles, which support cluster synchronization and allow a network to be reduced to a smaller quotient \cite{Xiao_symmetry_2008,SanchezGarcia_Symmetry_2020,nyberg2015mesoscopic,Sorrentino_2007,pecora_cluster_2014}. For continuous-time coupled-cell systems, the groupoid formalism shows how the network architecture constrains the admissible equations and their synchrony patterns \cite{golubitsky_nonlinear_2006,stewart_groupoids_2003}. Much less is known about symmetries of the global dynamics of discrete-time networks, that is, transformations of the state space that preserve both the dynamics and structure, or topology, of the network. Related questions have been studied through equivalences of finite dynamical systems \cite{laubenbacher_equivalence_2001}, symmetrizable networks \cite{aledo_symmetrizable_2023}, and interaction graphs of networks with isomorphic dynamics \cite{bridoux_interaction_2023}. Closest to the present work, Glass placed the state transition diagrams of networks without self-input on the edges of the $N$-dimensional hypercube and used its symmetry group, together with Burnside’s lemma, to count classes of equivalent networks~\cite{glass1975classification}. Fabre-Monplaisir et al.\ showed that transforming a Boolean network by a symmetry of the hypercube gives isomorphic synchronous and asynchronous dynamics and an isomorphic regulatory graph, and used the resulting classes to analyze regulatory motifs~\cite{fabre-monplaisir_isometries_2021}. In these works the hypercube symmetries are taken as given. Here we ask instead which state permutations preserve both the dynamics and the interaction topology of every BN, and show that for generic random-topology ensembles the hypercube symmetries are the complete answer. 

We identify the dynamical symmetries present in ensembles of the state-space networks (SSNs) of synchronous BNs. An SSN is a directed network that captures the complete dynamics of a BN: each of its nodes is a state of the BN, and each link a transition from one state to the next. An admissible symmetry operation transforms an SSN isomorphically into another SSN that corresponds to the dynamics of another BN with the same topology. We show that node-flips and node-label permutations generate a group $G_{F,P}$ that is admissible for every BN and is the complete symmetry group of generic random-topology ensembles. We give a test, based on topology alone, for the additional permutations allowed by a specific topology, and identify input-label permutations as a gauge freedom of the dynamics. The Zyklenzeiger group has been identified as the symmetry of the classes of BFs observed in evolutionary BN systems~\cite{bassler_evolution_2004,reichhardt_canalization_2007,matthew_d_reichl_canalization_2011}; see also Refs.~\cite{liu_finite_2011,shabnam_hossein_symmetry_2014}. Here we show that this symmetry follows from the global symmetry of the state-space dynamics together with this gauge freedom. We also count the symmetry orbits for any fixed topology using Burnside’s lemma, and study the effects of self-inputs and repeated inputs.

We formulate BN dynamics in a linear-algebraic language, in which states are basis vectors and the dynamics is a transition matrix that factors into a structure matrix and a logic matrix, $T = MS$~\cite{cheng2011analysis}. This formulation is what makes the symmetries transparent. A state permutation acts by conjugation, $T \rightarrow GTG^{-1}$. Admissibility is the condition that the conjugated matrix factors again with the same structure. The gauge freedom of input-label permutations is an ambiguity of this factorization. The Kronecker-product structure of $M$ shows directly how the global group $\GFP$ acts on the BF of each node, which is how the Zyklenzeiger symmetry arises, Eq.~\eqref{eq:local-action}. The same matrices give the orbit counts of Appendix~\ref{app:burnside}.

\section{Synchronous Boolean networks}
\label{sec:Synchronous BN}
We use a bra–ket notation borrowed from quantum mechanics, in which each Boolean state is a basis vector and the dynamics is a linear operator on the $2^N$-dimensional state space; in this language a symmetry is a similarity transformation of that operator, as in quantum mechanics~\cite{tinkham_group_1964}.

Consider a synchronous BN with \(N\) nodes. Each node $i$ has $K_i$ directed links that point to it from its in-neighbor nodes. The links describe its regulatory topological structure. At a time step $t$, each node $i$ has a Boolean state $\ket{x_i(t)}\in \{0,1\}$. The dynamics of a node $i$ is a Boolean function (BF) of the states of an ordered list $\{ i_1,\dots, i_{K_i}\}$ of the $K_i$ in-neighbor nodes of $i$, such that 
\begin{equation}
\ket{x_i(t+1)}=f_i\bigl(\hat{x}_i(t)\bigr)=\ket{f_i(\hat{x}_i)},
\label{eq:update}
\end{equation}
where 
\begin{equation}
\ket{\hat{x}_i(t)}\equiv\bigotimes_{j=1}^{K_i}\ket{{x_{i_j}}(t)} = \ket{x_{i_1}(t)\,\cdots\, x_{i_{K_i}}(t)}
\label{eq:u_i-def}
\end{equation}
is the \emph{input vector} for node $i$. Here, $\bigotimes$ indicates a Kronecker product and $\ket{\hat{x}_i(t)}\in \{0,1\}^{K_i}$. The state of every node is updated simultaneously. In general, each node has a different BF. For a node with in-degree $K$, there are $2^K$ possible input vectors, and since a BF must give a Boolean value for every possible input vectors there are $2^{2^K}$ possible BFs. 

At time $t$, the \emph{state vector} of the BN is the vector of the Boolean states of the nodes
\begin{equation}
\ket{x(t)} \equiv\bigotimes_{i=1}^N\ket{x_i(t)}= \ket{x_1(t)\,\cdots\, x_N(t)}.
\end{equation}
The SSN of a BN is a directed network of out-degree one, in which the nodes are the $2^N$ possible states of the network and the links indicate the transitions according to Eq.~{\eqref{eq:update}} from $\ket{x(t)}\rightarrow \ket{x(t+1)}$ for all $\ket{x(t)}\in \{0,1\}^N$. Self-links are possible in an SSN. In general, an SSN will break into multiple connected components. Each component will have one attractor, a cycle of length $l$. The number of states in a connected component is the size of its basin of attraction for the attractor in that connected component.

\begin{figure}[t]
\centering
\includegraphics{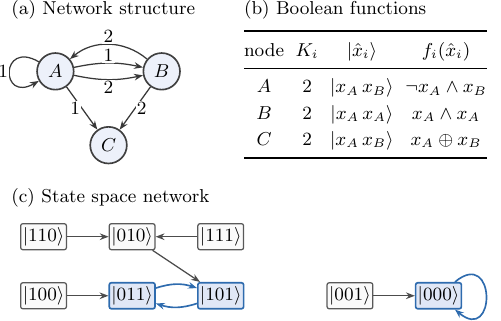}
\caption{Labeled structure, functions, and dynamics of an example three-node Boolean network. (a)~Regulatory structure of the BN. Labels specify each node and the order of the inputs for each node. (b)~In-degrees $K_i$, ordered inputs, and Boolean functions of each node{\color{red} }. 
(c)~The eight-state SSN, with states written \(\ket{x_A\,x_B\,x_C}\), contains the period-two attractor \(\ket{011}\leftrightarrow\ket{101}\) with basin size six and the fixed point \(\ket{000}\) with basin size two. Blue and black edges denote attractor and transient transitions, respectively.}
\label{fig:network}
\end{figure}

Figure~\ref{fig:network} shows a BN with three nodes. Node \(A\) has a self-input and has input vector \(\ket{\hat{x}_A}=\ket{x_A\,x_B}\), node \(B\) has a repeated input and has input vector \(\ket{\hat{x}_B}=\ket{x_A\,x_A}\), and node \(C\) has input vector \(\ket{\hat{x}_C}=\ket{x_A\,x_B}\). The BFs of the nodes are shown in the figure. ($\land$ indicates AND; $\lor$ indicates inclusive or; $\oplus$ indicates exclusive or; $\neg$ indicates negation.) The corresponding SSN contains a period-two attractor \(\ket{011}\leftrightarrow \ket{101}\), with a basin of six states, and a fixed-point attractor \(\ket{000}\), with a basin of two states. 
Define the \emph{transition matrix} $T$ of the SSN as
\begin{equation}
    T\equiv \sum_{x(t)\in\{0,1\}^N}\ket{x(t+1)}\bra{x(t)}.
\end{equation}
Then, the dynamics of the network is given by $\ket{x(t+1)}=T\ket{x(t)}$. Equivalently, \(T\) is the adjacency matrix of the SSN. It is a square matrix of size \(2^N\times 2^N\). It is a column-stochastic matrix; each state of the system has an out-degree one in the SSN, and thus each column of \(T\) contains a single 1, while the rest of its elements are 0.

A transition matrix for BN dynamics can be decomposed such that
\begin{equation}
T=MS,
\label{eq:TMS}
\end{equation}
where $S$ is the \emph{structure matrix} and $M$ is the \emph{logic matrix}~\cite{cheng2011analysis}. $S$ contains the information of the topological structure of the BN. $M$ contains the information of the dynamical behavior of each independent node. $S$, $M$, and $T$ are all static, time-invariant matrices.

Define the \emph{input vector} for the BN as the Kronecker product of the input vectors for each node
\begin{equation}
\ket{\hat{x}(t)}\equiv {\bigotimes_{i=1}^N}\ket{\hat{x}_i(t)}= \ket{\hat{x}_1(t)\,\cdots\,\hat{x}_N(t)},
\label{eq:u-def}
\end{equation}
and therefore has length $L=\sum_{i=1}^{N}K_i$, which is the number of links in the BN, with the input vector $\ket{\hat{x}(t)}\in \{0,1\}^{L}$. The structure matrix \(S\) is then defined as
\begin{equation}
S\equiv\sum_{x\in\{0,1\}^N}\ket{\hat{x}}\bra{x},
\label{eq:S-def}
\end{equation}
where the explicit time dependence in the notation is dropped because $S$ is time-independent. $S$ is a static matrix of size \(2^L \times 2^N\), in which the columns correspond to possible state vectors and the rows the possible input vectors. $S$ has one element of one in each column and the rest are zeros. The information in \(S\) is equivalent to that in the adjacency matrix of the BN.

The BF $f_i$ of a node $i$ can be represented as a matrix $M_i$, such that
\begin{equation}
M_i=\sum_{\hat{x}_i\in\{0,1\}^{K_i}}\ket{f_i(\hat{x}_i)}\bra{\hat{x}_i}.
\label{eq:Mi-def}
\end{equation}
$M_i$ is a \(2\times 2^{K_i}\) matrix in which the columns correspond to possible input vectors of the node and the rows its possible Boolean output vectors; each column contains a single 1 and a single 0. The logic matrix $M$ of the BN can be constructed as the Kronecker product
\begin{equation}
M\equiv\bigotimes_{i=1}^{N}M_i=\sum_{\hat{x}\in\{0,1\}^L}\ket{f(\hat{x})}\bra{\hat{x}},
\label{eq:M-def}
\end{equation}
where $\ket{f(\hat{x})}=\bigotimes_{i=1}^N\ket{f_i(\hat{x}_i)}$. $M$ is a static matrix of size \( 2^N \times2^L\). 

\section{Symmetry of State-space networks}
\label{sec:symmetry}
\subsection{Admissible transformations of $T$}

Consider the transition matrix $T$ of an SSN and the symmetry operation represented by a time-independent matrix $G$. The transformed transition matrix is
\begin{equation}
T' = G T G^{-1}.
\label{eq:conjugacy}
\end{equation}
Each of the $(2^N)!$ permutations of the $2^N$ states $\{\ket{x}\}$ is represented by a $2^N\times 2^N$ permutation matrix $G$ that by Eq.~\eqref{eq:conjugacy} will transform $T$ into $T'$. Thus, all the attractors, their periods, and the sizes of their basins of attraction are preserved, and so the SSNs of $T$ and $T'$ are isomorphic.

A symmetry transformation represented by $G$ is an \emph{admissible transformation} for a BN if the \(T'\) generated by it can be decomposed according to Eq.~\eqref{eq:TMS}  with a structure matrix that has the same topology as the original BN, that is, with the same pattern of regulatory interactions up to a relabeling of the nodes and order of their input links. 

Many of the $(2^N)!$ state permutations are not admissible transformations. For example, consider the two-node BN $x_1'=\neg x_1$ and $x_2'=x_1$, in which each node has in-degree one and both inputs come from node $1$. There are four states of the BN: $\ket{00}$, $\ket{01}$, $\ket{10}$, and $\ket{11}$. Each node has one input and four possible BFs, so there are $16$ possible BNs with this topology. If the states $\ket{00}$ and $\ket{10}$ are permuted, then no BN with the same topology can be constructed with the resulting $T'$. Thus, the permutation $\ket{00}\leftrightarrow\ket{10}$ is not an admissible transformation for this BN.

Two types of state permutations are admissible for every BN, and together they generate all state permutations with this property, which are therefore the ones applicable to ensembles of BNs. (See Appendix~\ref{app:GFP-admissible}.) The first type is node-flips \(F\), which flip the binary state of individual nodes while keeping the structure unchanged. The second type is node-label permutations \(P\), which keep the unlabeled topological structure of the BN unchanged. A third type of operation, input-label permutations \(R\), leaves $T$ unchanged but exploits a hidden gauge freedom in the decomposition of $T$ in Eq.~\eqref{eq:TMS}. $F$, $P$, and $R$ are all time-independent transformations. Thus, they transform the BN and its SSN for all time steps $t$. These three operations are considered in the following subsections. Additional admissible transformations can exist for BNs with a specific topological structure (Appendix~\ref{app:nonadmissible}).  None of these additional transformations are generally applicable to ensembles of BNs.

\begin{figure}[!t]
\centering
\includegraphics{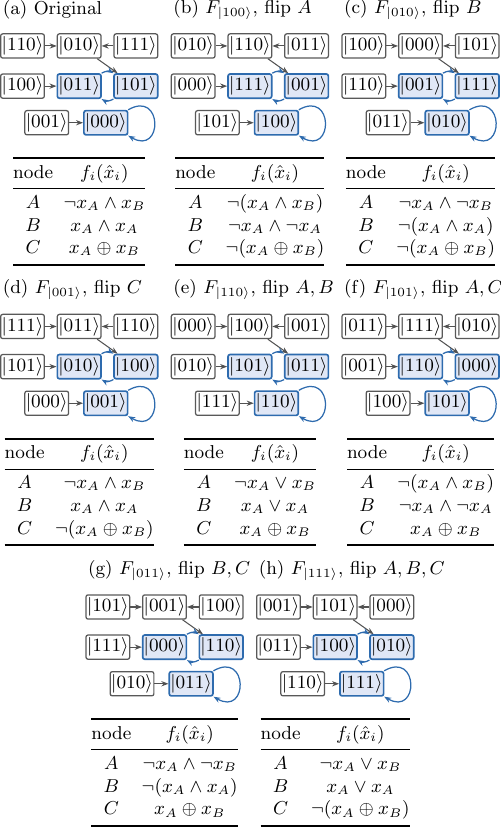}
\caption{Effects of node-flip operations $F$ on the example BN from Fig.~\ref{fig:network}.
(a)~SSN and BFs of the original BN in Fig.~\ref{fig:network}. (b)--(h)~Isomorphic SSNs \(T_b^{'}=F_{\ket{b}}TF_{\ket{b}}^{-1}\) and transformed BFs, obtained by flipping the node sets \(A\), \(B\), \(C\), \(AB\), \(AC\), \(BC\), and \(ABC\), respectively, i.e., with the flip vectors \(\ket{b}=\ket{100}\), \(\ket{010}\), \(\ket{001}\), \(\ket{110}\), \(\ket{101}\), \(\ket{011}\), and \(\ket{111}\). Each panel evaluates Eqs.~\eqref{eq:flip-local} and \eqref{eq:flip-conjugacy} for one flip vector. All transformations are applied independently to the original BN and preserve attractor periods and basin sizes. In all cases the structure matrix $S$ is unchanged. Blue and black edges denote
attractor and transient transitions, respectively.}
\label{fig:state-flip}
\end{figure}
\subsection{Node-flips $F$}
A node-flip changes the Boolean states of a subset of the nodes. The set of nodes to be flipped is specified by a binary \emph{flip vector}  $\ket{b}=\ket{b_1\,\cdots\, b_N}\in\{0,1\}^N$, where $\ket{b_j}=1$ if node $j$ is flipped and $\ket{b_j}=0$ if it is not. The operator for the flip of a single node is
\begin{equation}
F_{\ket{b_j}}=\sum_{x_j\in\{0,1\}}\ket{x_j\oplus b_j}\bra{x_j}.
\end{equation}
Thus, $F_1=\ket{1}\bra{0}+\ket{0}\bra{1}$ flips the node, and $F_0=\ket{0}\bra{0}+\ket{1}\bra{1}$ leaves it unchanged. The operator for the flip of all the nodes specified by $\ket{b}$ is
\begin{equation}
F_{\ket{b}}=\bigotimes_{j=1}^{N}F_{\ket{b_j}}=\sum_{x\in\{0,1\}^N}\ket{x\oplus b}\bra{x},
\end{equation}
where $\ket{x\oplus b}=\ket{x_1\oplus b_1\,\cdots\, x_N\oplus b_N}$, so that $F_{\ket{b}}\ket{x}=\ket{x\oplus b}$. Since flipping the same set of nodes twice restores every state, $F{_{\ket{b}}}^{-1}=F_{\ket{b}}$. $F_{\ket{b}}$ is a matrix of size $2^N \times 2^N$. 

Under a node-flip $F_{\ket{b}}$, the BF of each node can change because the flip can act on its input, on its output, or on both. The inputs of node $i$ are flipped according to the \emph{input flip vector} $\ket{\hat b_i}=\ket{b_{i_1}\,\cdots\,b_{i_{K_i}}}\in \{0,1\}^{K_i}.$
When a node-flip affects the BF of node $i$ at both its input and output, we have $\hat{x}_i' = \hat{x}_i \oplus \hat b_i$, and the output becomes $f_i(\hat{x}_i)\oplus b_i$. Thus, the transformed BF of node $i$ is
\begin{equation}
f_i'(\hat{x}_i')=f_i(\hat{x}_i'\oplus\hat b_i)\oplus b_i.
\label{eq:flip-local}
\end{equation}
The inputs of node $i$ are flipped by the operator
\begin{equation}
F_{\ket{\hat b_i}}=\bigotimes_{j=1}^{K_i}F_{\ket {b_{i_j}}}
=\sum_{\hat{x}_i\in\{0,1\}^{K_i}}\ket{\hat{x}_i\oplus\hat b_i}\bra{\hat{x}_i},
\label{eq:flip-Mi}
\end{equation}
which is a matrix of size $2^{K_i}\times 2^{K_i}$, and the transformed logic matrix of node $i$ is
\begin{equation}
M_i'=\sum_{\hat{x}_i\in\{0,1\}^{K_i}}\ket{f_i(\hat{x}_i)\oplus b_i}\bra{\hat{x}_i\oplus\hat b_i}
=F_{\ket{b_i}}M_iF_{\ket{\hat b_i}}.
\label{eq:flip-Mi-prime}
\end{equation}
The input flips of all nodes are collected in the operator
\begin{equation}
F_{\ket{\hat b}}=\bigotimes_{i=1}^{N}F_{\ket{\hat b_i}}=\sum_{\hat{x}\in\{0,1\}^{L}}\ket{\hat{x}\oplus\hat b}\bra{\hat{x}},
\label{eq:flip-hat}
\end{equation}
which is a matrix of size $2^L \times 2^L$. Since $M=\bigotimes_{i=1}^{N}M_i$, the mixed-product property of the Kronecker product gives
\begin{equation}
M'=F_{\ket{b}}MF_{\ket{\hat b}}
=\sum_{\hat{x}\in\{0,1\}^{L}}\ket{f(\hat{x})\oplus b}\bra{\hat{x}\oplus\hat b}.
\label{eq:flip-M}
\end{equation}

A flip does not change the structure matrix. Since $S$ only copies, reorders and discards node states, flipping the node states before applying $S$ is equivalent to flipping the corresponding input states afterwards. Thus, as proved in Proposition~\ref{prop:flip-S} of Appendix~\ref{app:GFP-admissible},
\begin{equation}
F_{\ket{\hat b}}S=SF_{\ket{b}}.
\label{eq:flip-S}
\end{equation}
Combining Eqs.~\eqref{eq:flip-M} and \eqref{eq:flip-S}, the transformed transition matrix is
\begin{equation}
T'=M'S=F_{\ket{b}}MF{_{\ket{\hat b}}}S=F_{\ket{b}}MSF_{\ket{b}}=F_{\ket{b}}T{F_{\ket{b}}}^{-1}.
\label{eq:flip-conjugacy}
\end{equation}
Thus, a flip changes only $M$, and $T'$ is related to $T$ by a conjugation of the form of Eq.~\eqref{eq:conjugacy}. Since $T'$ is generated by a BN with the same structure matrix $S$, node-flips are admissible transformations for every BN. Figure~\ref{fig:state-flip} shows all $2^N$ node-flips of the BN in Fig.~\ref{fig:network}, together with the transformed BFs. In each case, the transformed SSN is isomorphic to the original one.

\begin{figure}[!t]
\centering
\includegraphics{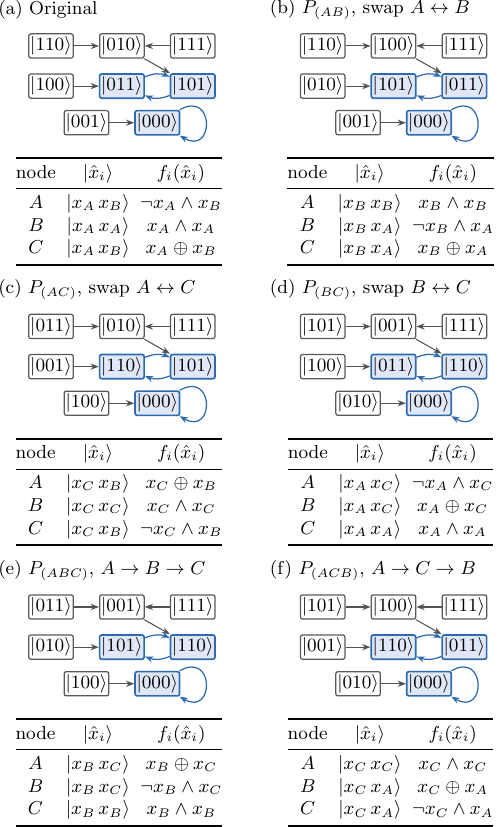}
\caption{Effects of node-label operations $P$ on the example BN from Fig.~\ref{fig:network}. (a)~SSN and BFs of the original BN in Fig.~\ref{fig:network}. (b)--(d)~Isomorphic SSNs obtained by exchanging node labels \(A\leftrightarrow B\), \(A\leftrightarrow C\), and \(B\leftrightarrow C\), respectively. (e)--(f)~Conjugate SSNs for the cyclic relabelings \(A\to B\to C\to A\), and its inverse \(A\to C\to B\to A\). Together, the panels show all six permutations. Each acts as \(T_{\pi}'=P_{\pi}T{P_{\pi}}^{-1}\), transporting the local
BFs and their input vectors together, as shown in the tables. Each panel evaluates Eqs.~\eqref{eq:sym-perm-local} and \eqref{eq:sym-node-conjugacy} for one permutation. States remain ordered \(\ket{x_A\,x_B\,x_C}\). All transformations are applied
independently to panel~(a) and preserve attractor periods and basin sizes. Blue and black edges denote attractor and transient transitions,
respectively.}
\label{fig:node-permutations}
\end{figure}

\subsection{Node-label permutations \texorpdfstring{\(P\)}{P}}
The second family of admissible transformations is node-label permutations, which rename the nodes. A node-label permutation $\pi$ is specified by a permutation mapping $\{i\rightarrow \pi(i); i=1, \dots, N\}$, which relabels the nodes in a new order. The BF of node $i$ becomes the BF of node $\pi(i)$. The unlabeled network topology and the set of BFs $\{f_i;i=1,\dots,N\}$ remain unchanged. The permuted nodes can be ordered so that $\{x_i'=x_{\pi^{-1}(i)};i=1,\dots, N\}$. Thus, the permuted state vector of the BN is 
\begin{equation}
    \ket{x'}=\bigotimes_{i=1}^N\ket{x'_i}.
\end{equation}
The operator for the permutation is
\begin{equation}
P_{\pi}=\sum_{x\in\{0,1\}^N}\ket{x'}\bra{x},
\label{eq:sym-node-perm}
\end{equation}
so that $P_{\pi}\ket{x}=\ket{x'}$. Unlike a flip, a permutation is generally not its own inverse; relabeling by the mapping $i\rightarrow\pi^{-1}(i)$ restores all labels, so ${P_{\pi}}^{-1}=P_{\pi^{-1}}$. 

The newly labeled node $i$ has $K'_i=K_{\pi^{-1}(i)}$ inputs and its input vector is 
\begin{equation}
    \ket{\hat{x}'_i}=\ket{\hat{x}_{\pi^{-1}(i)}}=\ket{x'_{i_{1}}\,\cdots\, x'_{i_{K'_i}}}.
\end{equation}
Thus, the BF and the logic matrix of the newly labeled node $i$ are
\begin{equation}
f_i'(\hat{x}_i')=f_{\pi^{-1}(i)}(\hat{x}_{\pi^{-1}(i)}), \quad\mbox{and}
\end{equation}
\begin{equation}
M_i'=M_{\pi^{-1}(i)}=\sum_{\hat{x}'_i\in\{0,1\}^{K'_i}}\ket {f'_i(\hat{x}'_i)}\bra{\hat{x}'_i}.
\label{eq:sym-perm-local}
\end{equation}
The permuted input vector for the BN is then
\begin{equation}
    \ket{\hat{x}'}=\bigotimes_{i=1}^N \ket{\hat{x}_i'},
\end{equation}
and the input permutation operator is
\begin{equation}
P_{\hat\pi}=\sum_{\hat{x}\in\{0,1\}^{L}}\ket{\hat{x}'} \bra{\hat{x}},
\label{eq:sym-P-hat}
\end{equation}
so that $P_{\hat\pi}\ket{\hat{x}}=\ket{\hat{x}'}$. $P_{\hat\pi}$ is a matrix of size $2^L\times2^L$. Finally, the transformed logic matrix is
\begin{equation}
M'=\bigotimes_{i=1}^{N}M'_i=\sum_{\hat{x}\in\{0,1\}^{L}}\ket{f'(\hat{x}')}\bra{\hat{x}'}=P_{\pi} M{P_{\hat\pi}}^{-1}.
\label{eq:sym-perm-M}
\end{equation}

Unlike a node-flip, a node-label permutation changes the structure matrix because both the nodes and their input nodes are relabeled. The transformed structure matrix is 
\begin{equation}
S'=P_{\hat\pi}S{P_{\pi}}^{-1}=\sum_{x'\in\{0,1\}^{N}}\ket{\hat{x}'}\bra{x'}.
\label{eq:sym-perm-S}
\end{equation}
$S'$ has the same topology structure as $S$. Combining Eqs.~\eqref{eq:sym-perm-M} and \eqref{eq:sym-perm-S}, the transformed transition matrix is
\begin{equation}
T'=M'S'=P_{\pi} M{P_{\hat\pi}}^{-1}P_{\hat\pi}S{P_{\pi}}^{-1}=P_{\pi} T{P_{\pi}}^{-1}.
\label{eq:sym-node-conjugacy}
\end{equation}
Since $S'$ is simply a relabeled copy of $S$ and $M'$ is constructed by permuting the BFs in $M$, node-label permutations $P_{\pi}$ are admissible transformations for every BN. Figure~\ref{fig:node-permutations} shows all possible node-label permutations of BN of Fig.~\ref{fig:network}. Again, in each case, the transformed SSN is isomorphic to the original one.

\subsection{Input-label permutations \texorpdfstring{\(R\)}{R} as a hidden gauge freedom}
\label{sec:L}

The decomposition of $T$ in Eq.~\eqref{eq:TMS} also admits a transformation that is not formally a symmetry operation but is, instead, a gauge freedom. This gauge freedom is input-label permutations, which reorder the inputs to one or more nodes. The order of the inputs to a node, which are its in-degree neighbors, is arbitrary. Thus, their order can be changed as long as the node’s BF is adjusted accordingly. An input-label permutation $\sigma_i$ of node $i$ maps its input labels as $\{i_j\rightarrow \sigma_i(i_j);\,j=1,\dots,K_i\}$. The permuted input states of node $i$ can then be ordered so that $\{x'_{i_j}=x_{\sigma_i^{-1}(i_j)};\,j=1,\dots,K_i\}$. Thus, the permuted input state vector of node $i$ is
\begin{equation}
    \ket{\hat{x}'_i}=\bigotimes_{j=1}^{K_i}\ket{x'_{i_j}}.
\end{equation}
The corresponding input permutation operator is
\begin{equation}
    R_{\sigma_i}=\sum_{\hat{x}_i\in\{0,1\}^{K_i}}\ket{\hat{x}'_i}\bra{\hat{x}_i},
    \label{eq:L}
\end{equation}
so that $R_{\sigma_i}\ket{\hat{x}_i}=\ket{\hat{x}'_i}$. $R_{\sigma_i}$ is a matrix of size $2^{K_i}\times 2^{K_i}$. Like node-label permutations, it is generally not its own inverse. Relabeling by mapping $i_j\rightarrow\sigma_i^{-1}(i_j)$ restores the original input order, so ${R_{\sigma_i}}^{-1}=R_{\sigma_i^{-1}}$.

To preserve the output of node $i$, the BF of the relabeled inputs must satisfy
\begin{equation}
    f'_i(\hat{x}'_i)=f_i(\hat{x}_i).
    \label{eq:input-local}
\end{equation}
Its logic matrix is therefore
\begin{align}
    M'_i &= \sum_{\hat{x}'_i\in\{0,1\}^{K_i}}\ket{f'_i(\hat{x}'_i)}\bra{\hat{x}'_i} \nonumber\\
          &= \sum_{\hat{x}'_i\in\{0,1\}^{K_i}}\ket{f_i(\hat{x}_i)}\bra{\hat{x}'_i}=M_i{R_{\sigma_i}}^{-1},
    \label{eq:input-Mi}
\end{align}
which satisfies $M'_i\ket{\hat{x}'_i}=M_i\ket{\hat{x}_i}$. 

The input labels of all of the nodes can be permuted similarly. In general, the permuted input vector of a BN from an input-label permutation $\sigma$ is
\begin{equation}
    \ket{\hat{x}'}=\bigotimes_{i=1}^{N}\ket{\hat{x}'_i},
\end{equation}
and the input permutation operator of the BN is
\begin{equation}
    R_{\sigma}=\bigotimes_{i=1}^{N}R_{\sigma_i}=\sum_{\hat{x}\in\{0,1\}^{L}}\ket{\hat{x}'}\bra{\hat{x}},
    \label{eq:input-R-net}
\end{equation}
so that $R_{\sigma}\ket{\hat{x}}=\ket{\hat{x}'}$. $R_{\sigma}$ is a matrix of size $2^L\times 2^L$. 

If the inputs in $S$ and the corresponding inputs of the BFs in $M$ are permuted together, the transformed structure and logic matrices become
\begin{align}
    S'&=R_{\sigma}S=\sum_{x\in\{0,1\}^N}\ket{\hat{x}'}\bra{x}, \nonumber\\
    M'&=\bigotimes_{i=1}^{N}M'_i=M{R_{\sigma}}^{-1}=\sum_{\hat{x}\in\{0,1\}^{L}}\ket{f(\hat{x})}\bra{\hat{x}'},
    \label{eq:input-refactorization}
\end{align}
The transformed transition matrix is
\begin{equation}
    T'=M'S'=M{R_{\sigma}}^{-1}R_{\sigma}S=MS=T.
    \label{eq:input-gauge}
\end{equation}
Thus, any input-label permutation $R_{\boldsymbol{\sigma}}$ leaves $T$ unchanged. Because $R_{\boldsymbol{\sigma}}$ acts on the input vector space rather than on the state space, it does not relate different SSNs. Rather, it is a gauge freedom of the decomposition that changes how a BN is represented without changing its dynamics.

\subsection{The universal dynamical symmetry group and its completeness for BN ensembles}
\label{sec:group}
Since node-flips and node-label permutations generate all state permutations that are admissible for every BN, they generate a symmetry group of every BN ensemble, $\GFP$. $\GFP$ is the $N$-dimensional hyperoctahedral group $\cB_N$, which is the semidirect product
\begin{equation}
\GFP=\cB_N=(\Z_2)^N\rtimes\Sym_N,\quad |\GFP|=2^NN!.
\label{eq:B_N}
\end{equation}
Here $(\Z_2)^N$ is the group of node-flips and $\Sym_N$ is the group of node-label permutations. 

By a BN ensemble we mean a collection of BNs with interaction topologies drawn at random, in which every BF consistent with a topology can occur, each BN weighted by its probability.
Every state permutation preserves the attractors and basins of an SSN, but only admissible ones map a BN onto a BN with the same wiring, as is needed for a transformation of an ensemble of networks. A state permutation is a symmetry of the ensemble only if it is admissible for every BN the ensemble contains, mapping all BNs with a given topology to the same relabeled copy of that topology (Appendix~\ref{app:GFP-minimal}). This ensures that the ensemble is mapped onto itself; whether the statistical weights of the ensemble are also invariant is a separate condition, discussed in Sec.~\ref{sec:local}. Node-flips and node-label permutations always qualify, so $\GFP$ is a symmetry of every ensemble. Whether it is the complete symmetry group depends on which topologies the ensemble includes. If even one included topology admits no state permutation outside $\GFP$, then $\GFP$ is complete. This is the case for any topology in which no two nodes read the same set of nodes and all nodes have the same number of distinct inputs (Appendix~\ref{app:GFP-minimal}), such as the directed ring (Theorem~\ref{thm:GFP-minimal}). Every ensemble with a fixed in-degree $K<N$ contains such topologies, so for these ensembles, and for any ensemble that includes them, $\GFP$ is the complete symmetry group. This is not guaranteed for every ensemble, however. If every node reads every node, including itself, then every possible dynamics is realized by some BN, and every state permutation is admissible. Completeness also assumes that every BF can occur. If the BFs are restricted, the argument of Appendix~\ref{app:nonadmissible} no longer applies.

\begin{figure}
    \centering
    \includegraphics{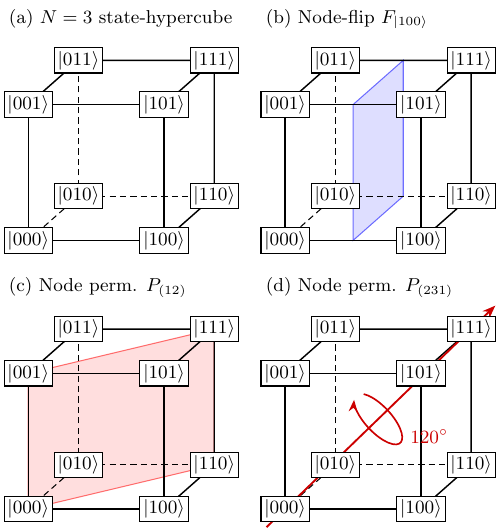}
    \caption{ Symmetry operations on the state-hypercube with $N=3$ nodes.
(a)~The three-dimensional hypercube has vertices that are the possible states of the BN $\bigl\{ \ket{x_1\,x_2\,x_3} \in \{0,1\}^3 \bigr\}$. (b)~Result of an example node-flip that flips the state of node $1$. It reflects the cube about the plane $x_1=1/2$ (blue). (c)~Result of a node-label permutation that permutes nodes
$1\leftrightarrow 2$. It reflects the cube about the plane $x_1=x_2$ (red). (d)~Result of a node-label permutation that cyclically permutes nodes
$1\rightarrow 2, 2\rightarrow 3, 3 \rightarrow 1$. It rotates the cube about the main diagonal line from $\ket{000}$ to $\ket{111}$ (red).}
\label{fig:hypercube}
\end{figure}

The group \(\GFP\) has a simple geometric interpretation. It is the symmetry group of the $N$-dimensional hypercube, whose $2^N$ vertices represent system states $\{\ket{x}\}$ and whose edges connect states that differ by flipping the state of a single node~\cite{Harrison1965, glass1975classification}. That these symmetries map the dynamics and the regulatory graph of a BN onto isomorphic ones is known~\cite{glass1975classification, fabre-monplaisir_isometries_2021}. What is new here is that they are the only state permutations admissible for every BN (Theorem~\ref{thm:GFP-minimal}), so that they form the complete symmetry group of generic BN ensembles. As shown in Fig.~\ref{fig:hypercube}, node-flips reflect the state-hypercube about hyperplanes that bisect edges, while node-label permutations rotate or reflect the state-hypercube about its main diagonal from the $\ket{00\dots0}$ vertex to the $\ket{11\dots1}$ vertex. 
\begin{figure}
\centering
\includegraphics{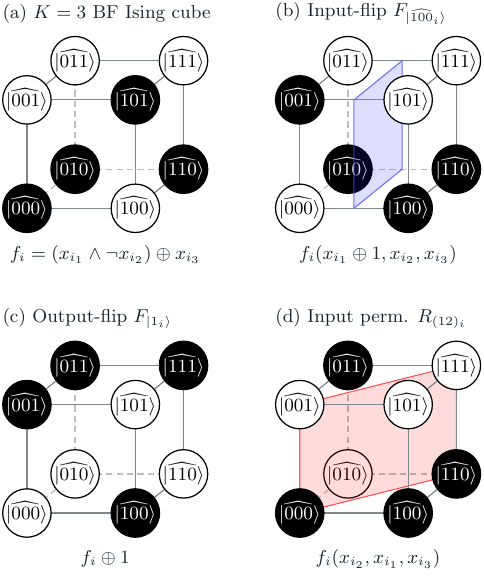} 
\caption {Symmetry operations on a  BF-hypercube with $K=3$ inputs.
(a) An example of a Boolean function for a node $i$, \(f_i=(x_{i_1}\land\neg x_{i_2})\oplus x_{i_3}\), represented as a three-dimensional hypercube with vertices that are possible input  vectors of node $i$, \(\bigl \{ \ket{x_{i_1}\,x_{i_2}\,x_{i_3}} \in \{0,1\}^3\bigr\}\). Each vertex is colored either black or white depending on whether the output of $f_i$ is $0$ or $1$, respectively, for the corresponding input vector. (b)~Result of an example node-flip \(F_{\ket{\hat{b}_i}}=F_{\ket{(\widehat{100})_i}}\) that flips the state of the first input node of node $i$. It reflects the cube about the plane \(x_{i_1}=1/2\) (blue). (c)~Result of an example node-flip \(F_{\ket{{b}_i}}=F_{\ket{1_i}}\) that flips the output of node $i$. It corresponds to a parity operation that changes the color of each vertex of the cube. (d)~Result of an example of an input-label permutation \(R_{{(12)}_i}\) that permutes the labels of the first two inputs of node $i$. It reflects the cube about the plane \(x_{i_1}=x_{i_2}\) (red). Panels (b)--(d) show the three factors of Eq.~\eqref{eq:local-action}, \(F_{\ket{\hat b_i}}\), \(F_{\ket{b_i}}\), and \(R_{\sigma_i}\), respectively.}
\label{fig:Z3-ising}
\end{figure}

Knowing the symmetry of SSNs allows us to combine isomorphic SSNs for BNs with a specific topology of regulatory interactions. Each orbit of $\GFP$ is a set of BNs whose SSNs map onto each other under elements of $\GFP$, so all members of an orbit have isomorphic SSNs. The converse need not hold. Two BNs with the same structure can have isomorphic SSNs without being related by $\GFP$, through a state permutation that is admissible only for that structure (Appendix~\ref{app:nonadmissible}) or only for those particular BNs. 
Such isomorphic dynamics that are not related by a hypercube symmetry were also noted in Refs.~\cite{glass1975classification,fabre-monplaisir_isometries_2021}. The orbits of $\GFP$ therefore refine the classes of BNs with isomorphic SSNs. For the directed ring with $N = 3$, the two coincide (Appendix~\ref{app:ring3}). The number of orbits of a group $G$ is given by Burnside's lemma
\begin{equation}
|X/G|=\frac{1}{|G|}\sum_{g\in G}|X^g|,
\label{eq:burnside}
\end{equation}
where $X$ is the set of BNs with the given structure, including its relabeled copies (the expanded ensemble of Appendix~\ref{app:burnside}), $g$ is a symmetry operation in $G$, and $X^g$ is the set of elements of $X$ left unchanged by $g$. The result depends on the structure of the BN through its automorphisms and how they act on the input states of its nodes, as shown in Appendix~\ref{app:burnside}.

\section{Local symmetry of Boolean functions}
\label{sec:local symmetry}

\subsection{Zyklenzeiger symmetry}
\label{sec:local}
The dynamical symmetry of ensembles of BNs leads to a local symmetry of BFs in those ensembles. Previous work studying the evolution of BN ensembles has found that, when the evolution has reached a steady state, BFs can be classified into sets by the frequency at which they occur in the BN ensembles. When the BFs have no self-inputs and no repeated inputs, these sets have been found to be orbits of the Zyklenzeiger group in large evolved networks~\cite{bassler_evolution_2004,reichhardt_canalization_2007,matthew_d_reichl_canalization_2011}; see also Refs.~\cite{liu_finite_2011,shabnam_hossein_symmetry_2014}. The Zyklenzeiger group is the direct product of the $K$-dimensional hyperoctahedral group $\cB_K$ and parity $\Z_2$
\begin{equation}
\cZ_K=\cB_K\times\Z_2,\quad |\cZ_K|=2^{K+1}K!.
\label{eq:ZK}
\end{equation}
The name refers to P\'olya's cycle index (Zyklenzeiger), which Harrison used to count the classes of BFs under this group~\cite{Polya1937,Harrison1963}. From Eqs.~\eqref{eq:flip-Mi-prime} and~ \eqref{eq:input-Mi}, the logic matrix of a node $i$ can generally be transformed as 
\begin{equation}
 M'_i=F_{\ket{b_i}}M_iF_{\ket{\hat b_{i}}}^{-1}{R_{\sigma_i}^{-1}},
\label{eq:local-action}
\end{equation}
where $F_{\ket{b_i}}$ is the node-flip of the output of node $i$, $F_{\ket{\hat b_{i}}}$ is a node-flip of the inputs of node $i$, and $R_{\sigma_i}$ is an input-label permutation of node $i$. If node $i$ has no self-input, $F_{\ket{b_i}}$ is independent of $F_{\ket{\hat b_{i}}}^{-1}$.

To understand the effects of symmetry transformations on a BF, or equivalently its logic matrix $M_i$, recognize that a BF of $K$ inputs can be represented as a two-coloring of the vertices of a $K$-dimensional hypercube, also known as an Ising hypercube~\cite{reichhardt_canalization_2007}. This Ising hypercube representation of a BF is distinct from the state-hypercube representation of a BN described in the previous section.

In a BF without self-inputs or repeated inputs, hyperoctahedral symmetry $\cB_K$ is generated differently than in a state-hypercube. In the BF of a node $i$, it is generated by node-flips of the input nodes of node $i$, which give the reflections $(\Z_2)^K$, and input-label permutations of node $i$, which give the permutations $\mathbb{S}_K$, rather than by node-flips and node-label permutations, as in the state hypercube. Node-label permutations do not act on the BF of a single node. As shown in Fig.~\ref{fig:Z3-ising}, node-flips of input nodes to a BF $F_{\ket{\hat{b}_i}}$ reflect the Ising hypercube about hyperplanes that bisect edges. A node-flip of the output of a BF $F_{\ket{b_i}}$ changes the color of each vertex of the Ising hypercube. This color change operation, also called a parity operation, has symmetry $\Z_2$ and commutes with all elements of $\cB_K$. Extending $\cB_K$ by a direct product with $\Z_2$ produces the Zyklenzeiger group $\cZ_K$, Eq.~\eqref{eq:ZK}. 

The Zyklenzeiger symmetry preserves the canalization properties of BFs. Canalization refers to the ability of a subset of inputs to a BF to control its output regardless of the state of other inputs~\cite{Waddington1942, Kauffman1993}. In a BF-hypercube representation, canalization occurs when a facet of the hypercube, i.e., a $(K-1)$-dimensional face on which one input has a fixed value, is uniformly colored~\cite{reichhardt_canalization_2007}. For $K=2$ this is an edge and for $K=3$ a square face. 

The Zyklenzeiger symmetry of BF frequencies follows from $\GFP$ under two conditions on the ensemble. First, the statistical weight of a BN must be unchanged by node-flips and node-label permutations. This holds, for example, when the distribution of topologies does not single out particular nodes and the weight of a BN otherwise depends only on its SSN up to isomorphism, i.e., on unlabeled properties such as the number, periods, and basin sizes of its attractors. Of node-flips and node-label permutations, only node-flips act on the BF of a given node;  invariance under node-label permutations ensures that the BF frequencies are the same at every node. Evolutionary games, discussed below, are a case in which this holds only approximately. Second, the order in which the inputs of a node are listed must carry no weight, so that input-label permutations, which leave the dynamics unchanged, also leave the weight unchanged. Now consider a node $i$ with $K$ distinct inputs, none of which is $i$ itself. The frequency of a given BF at node $i$ is the total weight of all BNs in which node $i$ has that BF. Flipping one of the input nodes of $i$ flips the corresponding input of its BF. It also changes the BFs of that input node and of the other nodes that read it, but these are summed over and do not affect the frequency at node $i$. Since the flip leaves the weights unchanged, a BF and the BF with that input flipped occur with the same frequency. In the same way, flipping node $i$ itself complements the output of its BF, and an input-label permutation reorders its inputs. These operations generate $\cZ_K$, so the BF frequencies are constant on each $\cZ_K$ orbit. 

These conditions are satisfied by the evolutionary game of competing nodes in which the Zyklenzeiger classes were observed~\cite{bassler_evolution_2004,reichhardt_canalization_2007,matthew_d_reichl_canalization_2011}. There, the initial networks and the replacement BFs are drawn from distributions that depend only on the homogeneity of each BF, the initial state is random, and a node loses when its state agrees most often with the majority state on the attractor. These rules are exactly invariant under node-label permutations, input-label permutations, and the simultaneous flip of all nodes. Flipping a single node is different. It complements the state of that node at every step of the attractor while changing the majority by at most one vote. A node that agreed with the majority on most steps of the attractor therefore agrees with it on few steps after the flip, and vice versa, up to corrections of order $1/N$. The rules of the game are thus not invariant under single-node flips, and the corresponding invariance of the statistical weights is an emergent, approximate property of the evolved steady state rather than an exact symmetry of the rules. It holds if, in the steady state, how often a node agrees with the majority is distributed symmetrically about one half of the steps of the attractor, as is the case if the state of a node is uncorrelated with the majority. This is expected for large $N$, where the balance between nodes in states 0 and 1 is achieved collectively rather than node by node~\cite{liu_finite_2011}, and it is consistent with the observed power-law decay of the splitting of BF classes with $N$~\cite{liu_finite_2011}. In small networks, selection instead favors input-inverting BFs, whose outputs are anticorrelated with the majority~\cite{liu_finite_2011}. This correlation breaks the single-node-flip symmetry, while the simultaneous flip of all nodes remains exact. In ensembles of random BNs constructed to be critical, in which each BF is biased toward 0 or 1 with equal probability, the conditions hold exactly, and the observed classes of BFs on active nodes are unions of $\cZ_K$ orbits~\cite{shabnam_hossein_symmetry_2014}.

If node $i$ has a self-input or a repeated input, a single node-flip acts on several inputs, or on an input and the output, at once, and only the reduced symmetry of Sec.~\ref{sec:reduce} remains. For a uniform random ensemble, in which every BF is equally likely, this symmetry is trivially satisfied. It becomes nontrivial when selection makes some BFs more common than others, as in evolved ensembles. Symmetry alone requires only that all BFs in an orbit share the same frequency. Which orbits are favored, and whether different orbits have different frequencies, is determined by the evolutionary dynamics.

\subsection{Reduced local symmetry for self-inputs and repeated inputs}
\label{sec:reduce}

\begin{figure}[t]
\centering
\includegraphics{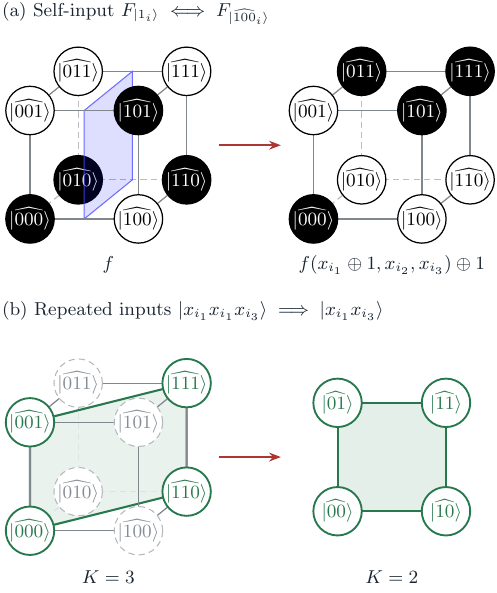}
\caption{
Examples of how a self-input and repeated inputs affect the $\cZ_3$ transformation. (a)~Self-input at $x_{i_1}$ with the BF in Fig.~\ref{fig:Z3-ising}(a). Flipping the state of node $i$ changes both its output
and its self-input, so the output-flip $F_{\ket{1_i}}$ and the input-flip
$F_{\ket{\widehat{100}_i}}$ occur together, as a reflection across the shaded
plane $x_1=1/2$ (blue) combined with a reversal of every color. (b)~Repeated inputs,
with inputs 1 and 2 from the same node. This reduces node \(i\)’s input state vector from \(\bigl\{\ket{x_{i_1}\,x_{i_1}\,x_{i_3}} \in \{0,1\}^3\bigr\}\) to \(\bigl\{\ket{x_{i_1}\,x_{i_3}} \in \{0,1\}^2\bigr\}\). Thus only four input states on the plane 
\(x_{i_1}=x_{i_2}\) (green) can occur, reducing the Ising cube to an Ising square.}
\label{fig:Z3-constrained}
\end{figure}

We have shown that Zyklenzeiger symmetry holds for BFs that do not have self-input or repeated inputs. However, if a node $i$ does have such inputs, then the Zyklenzeiger symmetry of the corresponding BF Ising hypercube is reduced or broken. First, consider a node \(i\) that has $K$ inputs and its BF $f_i$. Assume that one of its inputs is a self-input and the other \(K-1\) are distinct other nodes. Then, if the state of node $i$ changes as part of the node-flip operation $\ket{x_i}\rightarrow \ket{x_i \oplus 1}$, the input vector of node $i$ must also change. In this case, the hyperplane reflections of the Ising hypercube due to $F_{\ket{\widehat{b}_i}}$ and the parity operation on the Ising hypercube due to $F_{\ket{b_i}}$ are not independent. See Fig.~\ref{fig:Z3-constrained}(a). This reduces the symmetry group of the corresponding BF hypercube from $\cZ_K$ to $\cB_{K-1} \times \mathbb{Z}_2$, which is isomorphic to $\cZ_{K-1}$: the other $K-1$ inputs keep their hyperoctahedral symmetry, and the parity operation now flips the output and the self-input together. Here the self-input is listed first, so that input-label permutations act only on the other $K-1$ inputs. For $K=2$ and $3$, this group divides the 16 and 256 BFs into 7 and 34 orbits, compared with the 4 and 14 orbits of $\cZ_2$ and $\cZ_3$. 

Repeated inputs on node $i$ also reduce the Zyklenzeiger symmetry group $\cZ_K$ of its BF. For instance, if node $i$ with $K$ inputs has a  repeated input, so that two of the in-links come from the same other node, one input to the BF is redundant. This redundancy reduces the input state vector from $K$ dimensions to $K-1$ dimensions, so the Ising hypercube correspondingly becomes a $(K-1)$-dimensional Ising hypercube. See Fig.~\ref{fig:Z3-constrained}(b). This reduces the symmetry group of the corresponding BF hypercube from $\cZ_K$ to $\cZ_{K-1}$. Only the values of the BF on input vectors whose repeated inputs are equal affect the dynamics, so $\cZ_{K-1}$ acts on these effective BFs of $K-1$ inputs. For $K=3$, it divides the 16 effective BFs into 4 orbits. 

Repeated inputs and self-inputs change only the local symmetry of the BFs, not the dynamical symmetry group of the SSN $\GFP$. This is because $\GFP$ acts on the state space of the whole network and does not depend on how the BN is wired. The local group $\cZ_K$ is generated by the node-flips of  $\GFP$ together with input-label permutations. However, if there is a mechanism that breaks the global dynamical symmetry $\GFP$ of an ensemble of BNs, it will generally also break the local BF symmetry $\cZ_K$. An example is the evolutionary game at finite $N$~\cite{liu_finite_2011}. There, single-node flips are no longer symmetries of the ensemble, but node-label permutations and the simultaneous flip of all nodes remain exact. The simultaneous flip complements all inputs and the output of a BF, and together with input-label permutations it generates the subgroup $\mathbb{S}_K\times \Z_2$, of order $2\cdot K!$, consisting of input permutations and the simultaneous complement of all inputs and the output. For $K = 3$ this subgroup has 46 orbits, exactly the number of BF classes observed in finite networks, compared with the 14 orbits of $\cZ_3$ observed for large $N$~\cite{liu_finite_2011}.

\section{Summary and Discussion}
\label{sec:discussion}

In this paper, we studied the symmetry transformations that map the SSN of a BN onto an isomorphic SSN. Every state permutation maps the SSN of a BN onto an isomorphic SSN. However, a permutation is admissible only if the transformed SSN is generated by a BN with the same structure as the original BN, and only a subset of permutations satisfy this condition. The only transformations that are admissible for all possible network structures are the compositions of node-flips and node-label permutations, which form the symmetry group $\GFP$. 
Thus, for random BN ensembles, including every ensemble with a fixed in-degree $K < N$, these are the only admissible transformations (Sec.~\ref{sec:group}). For a specific topology, Theorem~\ref{thm:admissibility-test} decides from the topology alone which additional permutations are admissible. The group $\GFP$ is isomorphic to the $N$-dimensional hyperoctahedral group $\cB_N$ and the operations of the group can be interpreted geometrically as the symmetry operations of an $N$-dimensional state-space hypercube where each corner is a possible state of the BN.

Combined with the gauge freedom of input-label permutations, the global symmetry $\GFP$ produces the Zyklenzeiger symmetry $\cZ_K$ in the frequencies with which BFs occur in any BN ensemble whose statistical weights are invariant under $\GFP$, including large evolved ensembles, while the reduced symmetry of finite evolved networks reflects the subgroup of $\GFP$ that their dynamics respects (Sec.~\ref{sec:local symmetry}). $\cZ_K$ preserves the canalization properties of BFs (Sec.~\ref{sec:local})~\cite{reichhardt_canalization_2007}, so all BFs in an orbit share the same canalizing structure, and each set of BFs with a given canalizing structure is a union of orbits. Symmetry thus guarantees that BFs related by $\cZ_K$ occur with the same frequency. Which orbits are favored, such as the selection for canalizing BFs in large evolved networks~\cite{bassler_evolution_2004,reichhardt_canalization_2007, matthew_d_reichl_canalization_2011}, is determined by the evolutionary dynamics. 

The operations of $\cZ_K$ can be interpreted geometrically as the symmetry operations of a two-coloring of a $K$-dimensional BF-hypercube where each corner is a possible input vector of the BF and its vertices are colored black or white, indicating the output of the BF for the corresponding input vector. Although the classes had been identified as orbits of $\cZ_K$~\cite{reichhardt_canalization_2007}, why the evolutionary dynamics should have this symmetry had not been explained. Understanding the origin of the Zyklenzeiger symmetry of BFs in BN ensembles resolves this long-standing puzzle~\cite{bassler_evolution_2004,reichhardt_canalization_2007,liu_finite_2011,matthew_d_reichl_canalization_2011,shabnam_hossein_symmetry_2014}.

Furthermore, understanding symmetry $\GFP$ can reduce the complexity of the analysis of BN ensembles. $\GFP$ divides the BNs with a given structure into orbits whose members have isomorphic SSNs, and Burnside’s lemma gives the number of these orbits. Thus, any dynamical property that is invariant under $\GFP$ can be studied with a single representative per orbit instead of the full ensemble. The same reduction holds for evolution models whose fitness is invariant under $\GFP$. The present analysis applies specifically to synchronous Boolean networks, in which all nodes update simultaneously. Node-flips and node-label permutations also map asynchronous dynamics onto isomorphic dynamics~\cite{glass1975classification,fabre-monplaisir_isometries_2021}, so $\GFP$ remains a symmetry for asynchronous updating. Whether it is still the complete symmetry group, and how the state-space network changes, are open problems. The matrix formulation used here extends to stochastic updating, where $T$ becomes a stochastic matrix and symmetries still act by conjugation, and so provides a natural starting point for these problems. 
 

\appendix

\section{State permutations and admissibility}
\label{app:GFP-admissible}

In this appendix, we prove the results on admissibility used in Sec.~\ref{sec:symmetry}. Every state permutation gives an isomorphic SSN, every element of $\GFP$ is admissible for every BN, and no other state permutation has this property.

\subsection{State permutations give isomorphic SSNs}
\label{sec:all_permutations}

A permutation $g$ of the $2^N$ states is represented by the $2^N\times2^N$ permutation matrix
\begin{equation}
G=\sum_{x\in\{0,1\}^N}\ket{g(x)}\bra{x},\qquad G^{-1}=G^{\mathsf T}.
\label{eq:G_ket}
\end{equation}
Two SSNs are isomorphic if there is a bijection $g$ of the states such that $\ket{x}\to\ket{y}$ is a link in the original SSN if and only if $\ket{g(x)}\to\ket{g(y)}$ is a link in the transformed SSN.

\begin{theorem}
\label{thm:all_permutations}
Let $T$ be the transition matrix of a BN and $G$ any state permutation. Then $T'=GTG^{-1}$ is a transition matrix, and its SSN is isomorphic to  $T$ through $g$.
\end{theorem}

\begin{proof}
Since $G^{-1}=G^{\mathsf T}$, $\bra{x}G^{-1}=\bra{g(x)}$. Thus, Eq.~\eqref{eq:conjugacy} and the definition of $T$ give
\begin{equation}
\begin{aligned}
T'&=\sum_{x(t)}G\ket{x(t+1)}\bra{x(t)}G^{-1}\\
&=\sum_{x(t)}\ket{g(x(t+1))}\bra{g(x(t))}.
\end{aligned}
\label{eq:conjugate}
\end{equation}
Since $g$ is bijective, $\ket{g(x(t))}$ visits every state in $T'$ in the same way that $\ket{x(t)}$ visits every state in $T$. Thus, every column of $T'$ contains a single transition, and $\ket{g(x)}\to\ket{g(y)}$ is a transition in $T'$ if and only if $\ket{x}\to\ket{y}$ is a transition in $T$.
\end{proof}

\subsection{Node-flips commute with the structure matrix}
\label{sect:S commute F}

\begin{proposition}
\label{prop:flip-S}
For any structure matrix $S$ and any flip vector $\ket{b}$, $F_{\ket{\hat b}}S=SF_{\ket{b}}$.
\end{proposition}
\begin{proof}
Write $\ket{\hat x}$ for the input vector of Eq.~\eqref{eq:u-def} in the state $\ket{x}$, so that $S=\sum_{x}\ket{\hat x}\bra{x}$ according to Eq.~\eqref{eq:S-def}. According to Eq.~\eqref{eq:flip-hat}, $F_{\ket{\hat b}}S=\sum_{x}\ket{\hat x\oplus\hat b}\bra{x}$. Since $F_{\ket{b}}$ is symmetric and has its own inverse, $\bra{x}F_{\ket{b}}=\bra{x\oplus b}$, renaming the summation variable $x\to x\oplus b$ gives $SF_{\ket{b}}=\sum_{x}\ket{\widehat{x\oplus b}}\bra{x}$. The $j$th input of node $i$ in the state $\ket{x\oplus b}$ is $\ket{x_{i_j}\oplus b_{i_j}}$, so $\widehat{x\oplus b}=\hat x\oplus\hat b$, and the two sums are identical.
\end{proof}

\subsection{$\GFP$ is admissible for every structure}

Theorem~\ref{thm:GFP-admissible} is closely related to Propositions 3 and 4 of Ref.~\cite{fabre-monplaisir_isometries_2021}, which show that a hypercube symmetry maps the regulatory graph of a Boolean dynamical system onto an isomorphic graph. Here we prove it for the structure matrix $S$, which also records inputs that a BF does not use, self-inputs, repeated inputs, and the order of inputs, so that admissibility is defined for a fixed topology.

\begin{theorem}
\label{thm:GFP-admissible}
Let $T$ be the transition matrix of a BN with any structure matrix $S$ and any logic matrix $M$, and let $G=F_{\ket{b}}P_{\pi}\in\GFP$. Then
\begin{equation}
GTG^{-1}=M'S',
\label{eq:GFP-TMS}
\end{equation}
where $S'=P_{\hat\pi}S{P_{\pi}}^{-1}$ is the relabeled copy of $S$ of Eq.~\eqref{eq:sym-perm-S}, and $M'=\bigotimes_{i=1}^{N}M'_i$, with
\begin{equation}
M'_i=F_{\ket{b_i}}M_{\pi^{-1}(i)}F_{\ket{\hat b'_i}}.
\label{eq:GFP-Mi}
\end{equation}
Here, $\ket{\hat b'_i}$ is the input flip vector of node $i$ in $S'$. Thus, every element of $\GFP$ is admissible for every BN.
\end{theorem}

\begin{proof}
Since $\GFP=(\Z_2)^N\rtimes\Sym_N$, according to Eq.~\eqref{eq:B_N}, every element of $\GFP$ can be written as $F_{\ket{b}}P_{\pi}$. Since $G^{-1}={P_{\pi}}^{-1}F_{\ket{b}}$, inserting ${P_{\hat\pi}}^{-1}P_{\hat\pi}=I$ between $M$ and $S$ and using Eqs.~\eqref{eq:sym-perm-local}, \eqref{eq:sym-perm-M}, and \eqref{eq:sym-perm-S} gives
\begin{equation}
\begin{aligned}
GTG^{-1}
&=F_{\ket{b}}\bigl(P_{\pi}M{P_{\hat\pi}}^{-1}\bigr)\bigl(P_{\hat\pi}S{P_{\pi}}^{-1}\bigr)F_{\ket{b}}\\
&=F_{\ket{b}}\Bigl(\bigotimes_{i=1}^{N}M_{\pi^{-1}(i)}\Bigr)S'F_{\ket{b}}.
\end{aligned}
\label{eq:GFP-chain}
\end{equation}
Since Proposition~\ref{prop:flip-S} holds for every structure, $S'F_{\ket{b}}=F_{\ket{\hat b'}}S'$, where $F_{\ket{\hat b'}}=\bigotimes_{i=1}^{N}F_{\ket{\hat b'_i}}$ collects the input flips of $S'$. Substituting this into Eq.~\eqref{eq:GFP-chain} and using the mixed-product property of the Kronecker product gives Eq.~\eqref{eq:GFP-TMS}. Each $M'_i$ is a $2\times2^{K'_i}$ matrix with a single 1 in every column, so it is the logic matrix of a BF. Thus, $GTG^{-1}$ is generated by a BN with the structure $S'$.
\end{proof}

In the proof, no property of $S$ or $M$ was used, so Theorem~\ref{thm:GFP-admissible} holds also for structures with self-input or repeated inputs.

\subsection{$\GFP$ is the symmetry of BN ensembles}
\label{app:GFP-minimal}
For a structure $S$, let $\mathcal{A}(S)$ denote the set of state permutations that send every BN with the structure $S$ to a BN with the same relabeled copy $S'$ of $S$. A transformation with this property maps the set of BNs with the
structure $S$ as a whole onto the set of BNs with the structure $S'$, and so acts on an ensemble rather than on individual BNs. One could instead allow $S'$ to depend on the BN. Theorems~\ref{thm:GFP-admissible} and~\ref{thm:GFP-minimal} also hold with that weaker requirement, since their proofs do not use a common copy, but the two definitions differ in general. For $N=3$, let node 3 read node 2 and let nodes 1 and 2 have no inputs. The state permutation that exchanges the states of nodes 1 and 3 if and only if $x_2=1$ sends every BN with this structure to a BN on a relabeled copy, in which node 3 reads node 2 if $f_2=0$ and node 1 reads node 2 if $f_2=1$. No common copy exists, so this permutation is not in $\mathcal{A}(S)$. According to Theorem~\ref{thm:GFP-admissible}, $\GFP\subseteq\mathcal{A}(S)$ for every $S$. Here we show that equality holds for the directed ring BN, which is the structure of $N$ nodes in which the only input of node $i$ is node $i-1$, and the only input of node 1 is node $N$. 

\begin{theorem}
\label{thm:GFP-minimal}
For the directed ring, $\mathcal{A}(S)=\GFP$. Thus, the state permutations that are admissible for every BN are exactly the compositions of node-flips and node-label permutations, i.e .,
\begin{equation}
\bigcap_{S}\mathcal{A}(S)=\GFP,
\label{eq:GFP-minimal}
\end{equation}
where the intersection runs over all structures of $N$ nodes.
\end{theorem}

\begin{figure}[t]
\centering
\includegraphics{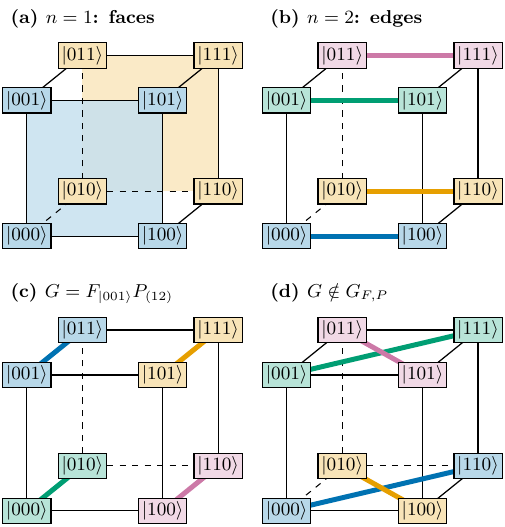}
\caption{Sets of states with the same $T\ket{x}$ for BNs on the directed ring with $N=3$, in which node 1 reads node 3, node 2 reads node 1, and node 3 reads node 2. States with the same color have the same $T\ket{x}$. (a)~$n=1$: $f_3=x_2$ and the other BFs are constant, so the sets are the faces $x_2=0$ and $x_2=1$. (b)~$n=2$: $f_2$ is constant and the other nodes have $f_i=x_{i_1}$, so the sets are the edges that connect states that differ only at node 1. (c)~$G=F_{\ket{001}}P_{(12)}\in\GFP$ maps these edges onto the edges that connect states that differ only at node 2. (d)~The state permutation $G\notin\GFP$ that flips node 2 if and only if $x_1=1$ maps them onto pairs of states that differ at nodes 1 and 2. These pairs are not connected by an edge, so they are not the sets of any BN on a directed ring.}
\label{fig:ring-proof}
\end{figure}

\begin{proof}
The proof uses the geometry of the $N$-dimensional state hypercube of Sec.~\ref{sec:group}. In the BN ring, every node reads a single node and is read by a single node. The BF of node $i$ is therefore either a constant function, $f_i=0$ or $f_i=1$, which gives the same output for both input states, or a non-constant function, $f_i=x_{i_1}$ or $f_i=\neg x_{i_1}$, which copies or flips its input, respectively. Thus, $T\ket{x}=T\ket{z}$ if and only if $\ket{x}$ and $\ket{z}$ agree at all nodes that are read by non-constant BFs. We call a set of states that a BN sends to the same successor a group. Since every node is read by a single node, a BN on a ring with $n$ non-constant BFs fixes the states of $n$ distinct nodes in each group, and leaves the other $N-n$ nodes free. It therefore has $2^n$ groups, each of which is an $(N-n)$-dimensional sub-cube of $2^{N-n}$ states, spanned by the edges along the $N-n$ free nodes. These sub-cubes are parallel to each other, i.e., they extend along the same $N-n$ directions, and together they partition the state hypercube. This holds wherever the non-constant BFs are located on the ring and also on every relabeled copy of the ring. For $N=3$, the groups are parallel faces for $n=1$ [Fig.~\ref{fig:ring-proof}(a)] and parallel edges for $n=2$ [Fig.~\ref{fig:ring-proof}(b)].

Now, let $G\in\mathcal{A}(S)$. Since $G$ only permutes states, the groups of $GTG^{-1}$ are the transformation under $G$ on $T$. Since $G$ is admissible, $GTG^{-1}$ is again a BN on a ring, so these images must again be parallel sub-cubes of the same dimension. In particular, consider $n=N-1$. The BN in which node $k+1$ is constant and every other node copies the node that it reads has as its groups the edges along node $k$ [Fig.~\ref{fig:ring-proof}(b)]. Thus, $G$ maps every edge of the state hypercube onto an edge, so it is an element of the symmetry group of the hypercube, which is the hyperoctahedral group $\GFP$. Together with Theorem~\ref{thm:GFP-admissible}, this proves the theorem.
\end{proof}

Figures~\ref{fig:ring-proof}(c) and \ref{fig:ring-proof}(d) compare an element of $\GFP$ with a state permutation outside $\GFP$, which maps edges onto pairs of states that are not connected by an edge.

The same conclusion holds for other structures. By Theorem~\ref{thm:admissibility-test}, whether a state permutation is admissible for every BN with a given structure depends only on which nodes read which, not on how those inputs are ordered or repeated. Thus, a ring in which each node reads its predecessor through several repeated inputs admits exactly the same state permutations as the ring itself.

More generally, only $\GFP$ is admissible for any structure in which no two nodes read the same set of nodes and every node reads the same number $K$ of distinct nodes. For unequal in-degrees, it is sufficient that any two nodes together read more distinct nodes than the largest number read by a single node. Let $g$ be a state permutation outside $\GFP$. As shown in Appendix~\ref{app:dependency}, $g$ then mixes the states of different nodes, i.e., for some node $i$, the new state of $i$ under $g$ depends on at least two distinct nodes, $j_1$ and $j_2$. Since their input sets differ, $j_1$ and $j_2$ together read at least $K+1$ distinct nodes. Since $g^{-1}$ is a bijection, any $m$ of its output states take all $2^m$ values, so together they must depend on at least $m$ nodes. Thus, the states $g^{-1}(y)_k$ of these $K+1$ nodes depend together on at least $K+1$ nodes. Following the chain $n\to k\to j\to i$ of Appendix~\ref{app:test}, row $i$ of the left-hand side of Eq.~(\ref{eq:admissibility-test}) therefore contains at least $K+1$ ones, whereas in every relabeled copy of the structure node $i$ reads only $K$ nodes, so row $i$ of $\hat{A}'$ contains only $K$ ones. By Theorem~\ref{thm:admissibility-test}, $g$ fails the test for every $S'\in\cS$, so $g\notin \mathcal{A}(S)$. Examples are the directed ring, the fully connected network without self-inputs, and the structure in which node $i$ reads nodes $i-1,\dots,i-K$, with indices taken modulo $N$, which exists for every $K<N$.
For $K = 1$ the structure is the directed ring, so this argument gives a second, algebraic proof of Theorem~\ref{thm:GFP-minimal}.

The condition fails if every node reads every node, including itself. Then every map of the state space onto itself is the dynamics of some BN, and every state permutation is admissible.

\section{Admissible transformations outside \texorpdfstring{$\GFP$}{G\_FP}}
\label{app:nonadmissible}

For a specific structure, a state permutation outside $\GFP$ can be admissible for every BN. In this appendix, we derive a test of admissibility that involves only the structure and no BF.

A state permutation $g$ is admissible for a BN with the transition matrix $T$ if $T'=GTG^{-1}$ is generated by a BN with the structure $S$ or with one of its relabeled copies $S'=P_{\hat\pi}S{P_{\pi}}^{-1}$ of Eq.~\eqref{eq:sym-perm-S}. Since the new state of a node depends only on its input vector, $T'$ is generated by a BN with the structure $S$ if and only if, for every node $i$,
\begin{equation}
\ket{\hat x_i}=\ket{\hat z_i}\quad\Longrightarrow\quad y_i=w_i,
\label{eq:admissible}
\end{equation}
where $\ket{y}=T'\ket{x}$ and $\ket{w}=T'\ket{z}$. That is, two states with the same input vector of node $i$ must give the same new state of node $i$. Checking Eq.~\eqref{eq:admissible} for every BN would require all $\prod_{i}2^{2^{K_i}}$ logic matrices. Instead, we compare which nodes the structure allows each node to read with which nodes a map actually uses.

\subsection{The interaction matrix}
\label{app:dependency}

The test of Eq.~\eqref{eq:admissible} must be applied to every BN separately. In this appendix, we derive a test that depends only on the state permutation $g$ and on the inputs allowed by the structure, and that decides whether $g$ is admissible for every BN with that structure. The test is based on the interaction graph of a map, which records the nodes on which the new state of each node actually depends.

The inputs allowed by a structure $S$ are recorded by its input pattern $\hat A$, i.e., the $N\times N$ matrix with $\hat A_{ij}=1$ if node $j$ is one of the inputs $i_1,\dots,i_{K_i}$ of node $i$, and $\hat A_{ij}=0$ otherwise. A self-input gives $\hat A_{ii}=1$, and a repeated input gives a single 1. The relabeled copy $S'$ has the input pattern $\hat A^{\pi}$, with $\hat A^{\pi}_{\pi(i)\pi(j)}=\hat A_{ij}$.

The inputs actually used by a map are recorded by its interaction graph~\cite{gadouleau_influence_2020}. For a map $h:\{0,1\}^N\to\{0,1\}^N$, we write $h(x)_i$ for the state of node $i$ in $\ket{h(x)}$, and $\ket{x\oplus e_j}$ for the state that differs from $\ket{x}$ only at node $j$. The new state of node $i$ depends essentially on node $j$ if changing only $x_j$ can change it. We denote by $B(h)$ the $N\times N$ interaction matrix with
\begin{equation}
B(h)_{ij}=1\Longleftrightarrow h(x)_i\neq h(x\oplus e_j)_i\ \text{for some state}\ \ket{x},
\label{eq:dependency}
\end{equation}
and $B(h)_{ij}=0$ otherwise. Robert called $B(h)$ the incidence matrix of $h$~\cite{robert_discrete_1986}, and it is the transpose of the adjacency matrix of the interaction graph in the convention of Ref.~\cite{gadouleau_influence_2020}. For a matrix $H=\sum_{x}\ket{h(x)}\bra{x}$, such as $G$, $F_{\ket{b}}$, $P_\pi$ or $T$, we write $B(H)=B(h)$.

Since $T\ket{x}=\ket{f(\hat x)}$, $B(T)$ records the inputs that the BFs of a BN actually use. A BN with the structure $S$ can only use the inputs of $S$, so $B(T)\le\hat A$, where $\le$ holds entrywise. Conversely, if $B(T)\le\hat A$, node $i$ can compute its new state from one input slot of each node $j$ with $\hat A_{ij}=1$ and ignore the other slots, so $T$ is generated by a BN with the structure $S$. Thus, the set $\mathcal{T}(S)$ of transition matrices of BNs with the structure $S$ is the set of maps whose interaction graph is contained in the input pattern of $S$, which is the class of maps studied in Ref.~\cite{gadouleau_influence_2020}. In particular, $\mathcal{T}(S)$ depends on $S$ only through $\hat A$, also when $S$ contains self-inputs or repeated inputs.

Interaction matrices compose through the Boolean product $\odot$, which is the ordinary matrix product with every positive entry replaced by 1~\cite{robert_discrete_1986},
\begin{equation}
(Q\odot Q')_{ij}=1\Longleftrightarrow Q_{ik}=Q'_{kj}=1\ \text{for some}\ k.
\label{eq:boolean-product}
\end{equation}
The Boolean product is associative and monotone in both factors. For two maps $u$ and $v$,
\begin{equation}
B(u\circ v)\le B(u)\odot B(v),
\label{eq:chain-rule}
\end{equation}
since a change of $x_n$ changes only the states $v(x)_k$ with $B(v)_{kn}=1$, and changing these one at a time shows that $u(\cdot)_i$ changes with at least one of them. Equality need not hold, since the dependencies along two different chains can cancel.

The interaction matrix identifies the elements of $\GFP$. A node-flip complements each node using only that node, so $B(F_{\ket{b}})=I$. A node-label permutation moves the state of node $j$ to node $\pi(j)$, so $B(P_\pi)=\Pi_\pi$, where $\Pi_\pi$ is the $N\times N$ permutation matrix with $(\Pi_\pi)_{ij}=1$ if $i=\pi(j)$. Thus, $B(g)=\Pi_\pi$ and $B(g^{-1})=\Pi_\pi^{\mathsf T}$ for $G=F_{\ket{b}}P_\pi$, and $\hat A^{\pi}=\Pi_\pi\odot\hat A\odot\Pi_\pi^{\mathsf T}$. Conversely, a bijection has no constant $g(x)_i$, so every row of $B(g)$ contains at least one 1. If every row contains exactly one 1, each $g(x)_i$ is the state of a single node or its complement, and since $g$ is a bijection, these nodes are distinct. Thus, $g(x)_i=x_{\pi^{-1}(i)}\oplus b_i$ for a permutation $\pi$, and $G=F_{\ket{b}}P_{\pi}\in\GFP$ according to Eq.~\eqref{eq:sym-node-perm}. Therefore, a state permutation outside $\GFP$ mixes the states of different nodes, i.e., some row of $B(g)$ contains two or more 1s.

The interaction graphs of BNs related by state permutations have been studied in Ref.~\cite{bridoux_interaction_2023}, which considers all state permutations $g$ for a fixed BN. Here, in contrast, we ask which $g$ are admissible for every BN with a given structure.

\subsection{The admissibility test}
\label{app:test}

One step of $T'=GTG^{-1}$, read from right to left, consists of three steps: $g^{-1}$ undoes the relabeling, the BN updates the state, and $g$ relabels it again. The new state of node $i$ under $T'$ can therefore depend on the old state of node $n$ only through a chain $n\to k\to j\to i$, in which $g^{-1}$ makes node $k$ use node $n$, node $k$ is an input of node $j$ in $S$, and $g$ makes node $i$ use node $j$. The Boolean product $B(g)\odot\hat A\odot B(g^{-1})$ records whether such a chain exists.

\begin{theorem}
\label{thm:admissibility-test}
Let $S'$ be $S$ or one of its relabeled copies, with the input pattern $\hat A'$. A state permutation $g$ sends every BN with the structure $S$ to a BN with the structure $S'$ if and only if
\begin{equation}
B(g)\odot\hat A\odot B(g^{-1})\le\hat A'.
\label{eq:admissibility-test}
\end{equation}
In this case, Eq.~\eqref{eq:admissibility-test} holds with equality.
\end{theorem}

\begin{proof}
The test is sufficient because of Eq.~\eqref{eq:chain-rule}. For every BN with the structure $S$, $B(T)\le\hat A$, so
\begin{equation}
\begin{split} 
B(GTG^{-1})\le B(g)\odot B(T)\odot B(g^{-1})\\
\le B(g)\odot\hat A\odot B(g^{-1})\le\hat A',
\end{split} 
\end{equation}
and $GTG^{-1}\in\mathcal{T}(S')$.

To show that the test is necessary, we show that every 1 of the left-hand side of Eq.~\eqref{eq:admissibility-test} is realized by some BN. Suppose that the left-hand side has a 1 at $(i,n)$, so that there are nodes $j$ and $k$ with $B(g)_{ij}=\hat A_{jk}=B(g^{-1})_{kn}=1$. Since $B(g)_{ij}=1$, there is a state $\ket{z}$ such that changing only $z_j$ changes $g(z)_i$. Consider the BN with the structure $S$ in which node $j$ copies the input slot that reads node $k$, and every other node $r$ is the constant $z_r$. In this BN, the successor of $\ket{x}$ is $\ket{z}$ or $\ket{z\oplus e_j}$, depending on $x_k$, so $g(f(\hat x))_i=x_k\oplus c$, with a constant $c$. Thus, under $GTG^{-1}$, the new state of node $i$ in the state $\ket{y}$ is $g^{-1}(y)_k\oplus c$, which depends on $y_n$ since $B(g^{-1})_{kn}=1$. Admissibility therefore requires $\hat A'_{in}=1$.

Finally, the two arguments show that the left-hand side of Eq.~\eqref{eq:admissibility-test} is the entrywise OR of $B(GTG^{-1})$ over all BNs with the structure $S$. If $g$ is admissible, $T\mapsto GTG^{-1}$ maps $\mathcal{T}(S)$ onto $\mathcal{T}(S')$, since it is injective and the two sets have the same size, $\prod_{i}2^{2^{d_i}}$, where $d_i$ is the number of distinct inputs of node $i$. Thus, this OR is that of $B(T')$ over $\mathcal{T}(S')$, which is $\hat A'$.
\end{proof}
\begin{proposition}\label{prop:AS}
Let $\mathcal{H}(S)$ be the set of state permutations that pass the test of Eq.~(\ref{eq:admissibility-test}) with $\hat{A}'=\hat{A}$, i.e., that send every BN with the structure $S$ to a BN with the same structure $S$. Then $\mathcal{H}(S)$ is a group, and
$\mathcal{A}(S)=\{P_\pi h:\ \pi\in\mathbb{S}_N,\ h\in\mathcal{H}(S)\}.$
\end{proposition}

\begin{proof}If $h_1$ and $h_2$ send $\mathcal{T}(S)$ onto itself, so do $h_1h_2$ and $h_1^{-1}$, so $\mathcal{H}(S)$ is a group. If $g\in\mathcal{A}(S)$ sends $\mathcal{T}(S)$ onto $\mathcal{T}(S')$, where $S'$ is the copy of $S$ relabeled by $\pi$, then by Theorem~\ref{thm:GFP-admissible} $P_\pi^{-1}$ sends $\mathcal{T}(S')$ back onto $\mathcal{T}(S)$, so $h=P_\pi^{-1}g\in\mathcal{H}(S)$ and $g=P_\pi h$. Conversely, $P_\pi h$ sends $\mathcal{T}(S)$ onto $\mathcal{T}(S')$ by Theorem~\ref{thm:GFP-admissible}, so $P_\pi h\in\mathcal{A}(S)$.\end{proof}

Theorem ~\ref{thm:admissibility-test} is a counterpart of Ref.~\cite{bridoux_interaction_2023} with the roles exchanged: there a BN is fixed and $g$ ranges over all state permutations, whereas here the structure is fixed and the BN ranges over all BNs with that structure. The sufficiency half follows from Eq.~\eqref{eq:chain-rule}; the necessity half and the equality are, to our knowledge, new. 

Theorem~\ref{thm:admissibility-test} applies to a fixed target structure $S'$. Thus $\mathcal{A}(S)$ is exactly the set of state permutations that pass the test of Eq.~(\ref{eq:admissibility-test}) for some $S'\in\cS$, and by Proposition~\ref{prop:AS} it suffices to apply the test with $S'=S$. A 1 of the left-hand side of Eq.~\eqref{eq:admissibility-test} where $\hat A'$ has a 0 names the forbidden link, i.e., in $GTG^{-1}$, node $i$ would have to read node $n$. The test depends on $g$ only through the pair $\bigl(B(g),B(g^{-1})\bigr)$, so the node-flips, which do not change this pair, do not change the result. For $G=F_{\ket{b}}P_\pi\in\GFP$, Eq.~\eqref{eq:admissibility-test} gives $\Pi_\pi\odot\hat A\odot\Pi_\pi^{\mathsf T}=\hat A^{\pi}$, in agreement with Theorem~\ref{thm:GFP-admissible}.

Note that the necessity argument uses BNs with copying and constant BFs. In an ensemble without them, e.g., one in which every BF depends on all of its inputs, Eq.~\eqref{eq:admissibility-test} still guarantees that $GTG^{-1}$ has the structure $S'$, but it can be stricter than necessary, and $GTG^{-1}$ need not belong to the same ensemble.

\section{Symmetry orbits}
\label{app:burnside}

According to Sec.~\ref{sec:group}, the orbits of $\GFP$ are classes of BNs whose SSNs are isomorphic through elements of $\GFP$ and Burnside's lemma, Eq.~\eqref{eq:burnside}, gives their number. 
Glass used Burnside’s lemma with the hypercube symmetry group to count classes of networks without self-input~\cite{glass1975classification}. Here we evaluate it for an arbitrary fixed structure.

We first specify the set $X$ on which $\GFP$ acts. A node-label permutation maps $S$ onto a relabeled copy $S'$, which in general differs from $S$, so the set must contain all relabeled copies of $S$. We denote by $\cS$ the set of distinct relabeled copies of $S$, including $S$ itself. We call each pair $(M,S')$, with $S'\in\cS$ and $M$ a logic matrix on $S'$, a representation of a BN, and we call the set of all representations the expanded ensemble $\cE$.

The node-label permutations that map $S$ onto itself form its automorphism group $\Aut(S)$. Each copy of $S$ is produced by exactly $|\Aut(S)|$ of the $N!$ node-label permutations, so $|\cS|=N!/|\Aut(S)|$. Every copy has $|\cM|=\prod_{i=1}^{N}2^{2^{K_i}}$ logic matrices, since a relabeling only reorders the factors. Thus,
\begin{equation}
|\cE|=\frac{N!}{|\Aut(S)|}\,|\cM|.
\label{eq:expanded-ensemble}
\end{equation}
Taking $X=\cE$ in Eq.~\eqref{eq:burnside}, with $|\GFP|=2^NN!$ according to Eq.~\eqref{eq:B_N}, gives
\begin{equation}
|\cE/\GFP|=\frac{1}{2^NN!}\sum_{g\in\GFP}|\cE^g|,
\label{eq:burnside-E}
\end{equation}
where $\cE^g$ is the set of representations in $\cE$ that $g$ leaves unchanged.

\subsection{Reduction to the automorphisms of \texorpdfstring{$S$}{S}}
\label{app:orbit-reduced}

Most elements of $\GFP$ leave no representation unchanged. An element $g=F_{\ket{b}}P_{\pi}$ changes the structure of $(M,S')$ unless $\pi$ maps $S'$ onto itself. Furthermore, the copies of $S$ are related by relabelings, so they all contribute equally to Eq.~\eqref{eq:burnside-E}. It is therefore sufficient to consider $S$ itself with the $2^N|\Aut(S)|$ elements with $\pi\in\Aut(S)$, and to multiply by the number $N!/|\Aut(S)|$ of copies. Thus, Eq.~\eqref{eq:burnside-E} reduces to
\begin{equation}
|\cE/\GFP|=\frac{1}{2^N|\Aut(S)|}\sum_{\pi\in\Aut(S)}\ \sum_{b\in\{0,1\}^N}\bigl|\cM^{F_{\ket{b}}P_{\pi}}\bigr|,
\label{eq:orbit-count-S}
\end{equation}
where $\cM^{g}$ is the set of logic matrices on $S$ that $g$ leaves unchanged. Note that $\pi\in\Aut(S)$ may map $S$ onto $R_{\sigma}S$, i.e., onto $S$ with its input links reordered. In this case, the gauge freedom of Eq.~\eqref{eq:input-gauge} brings the transformed logic matrix back onto $S$.

\subsection{Counting by cycles of input states}
\label{app:orbit-cycles}

The size of $\cM^{g}$ follows from how $g$ moves the input states. According to Eq.~\eqref{eq:GFP-Mi}, $g$ moves each input state of a node $j$ to an input state of node $\pi(j)$, and it complements the output if node $\pi(j)$ is flipped. Repeated application of $g$ eventually returns every input state to where it started, so the input states of all nodes fall into cycles. We denote the number of these cycles by $C(g)$. A logic matrix is left unchanged if and only if, along each cycle, the output at each input state equals the output at the previous one, complemented if a flip occurs in between. Thus, the output can be chosen freely at one input state of each cycle, and this choice fixes the outputs on the rest of the cycle.

We call a cycle odd if the output is complemented an odd number of times around it, and even otherwise. Around an odd cycle, the output would have to equal its own complement, so no logic matrix is left unchanged. Therefore,
\begin{equation}
\bigl|\cM^{g}\bigr|=
\begin{cases}
2^{C(g)}, & \text{if every cycle is even},\\
0, & \text{otherwise}.
\end{cases}
\label{eq:fix-cycles}
\end{equation}
For example, for a single node that reads itself, the node-flip exchanges its two input states and complements the output at each step. This gives one even cycle, so $2^1=2$ BFs are left unchanged, the identity and the negation. In the Ising-hypercube picture of Sec.~\ref{sec:local}, each cycle is a closed loop of constraints between spins, with an antiferromagnetic bond wherever the output is complemented, and an odd cycle is frustrated.

\subsection{Example: the directed ring with \texorpdfstring{$N=3$}{N=3}}
\label{app:ring3}
As an example, we consider the directed ring of Fig.~\ref{fig:ring-proof}, in which node $j$ reads node $j-1$, with indices taken modulo 3. Every node has $K=1$ input and one of four BFs, the constants 0 and 1, the identity, and the negation, so $|\cM|=4^3=64$. The node-label permutations that map the ring onto itself are the three rotations $\tau^k$, $k=0,1,2$, where $\tau(j)=j+1$. Thus, $|\Aut(S)|=3$, and the only other copy of $S$ is the ring with the opposite orientation. According to Eq.~\eqref{eq:expanded-ensemble}, the expanded ensemble contains $|\cE|=(3!/3)\times64=128$ representations, and Eq.~\eqref{eq:orbit-count-S} involves the $2^3\cdot3=24$ elements $F_{\ket{b}}P_{\tau^k}$.

For $\pi=\tau^k$, the input of node $\pi(j)$ is node $\pi(j-1)$. Thus, according to Eq.~\eqref{eq:GFP-Mi}, $g=F_{\ket{b}}P_{\tau^k}$ moves the input state $u\in\{0,1\}$ of node $j$ to the input state $u\oplus b_{\pi(j)-1}$ of node $\pi(j)$, and it complements the output if $b_{\pi(j)}=1$.

For $k=0$, $g$ moves the input states of each node $j$ among themselves. If $b_{j-1}=0$, both input states stay in place and form two cycles of length one. These cycles are odd if $b_j=1$, since the output is then complemented once. If $b_{j-1}=1$, the two input states are exchanged and form a single cycle of length two. Along it, the output is complemented $2b_j$ times, so it is even. Therefore, $|\cM^g|\neq0$ only if every flipped node reads a flipped node, which on the ring allows only $b=000$ and $b=111$. For $b=000$, $g$ is the identity, $C(g)=6$, and all $2^6=64$ logic matrices are left unchanged. For $b=111$, $C(g)=3$, and the $2^3=8$ logic matrices that are left unchanged are those in which every node has the identity or the negation, as in the example of a single node that reads itself.

For $k=1$ and $k=2$, $\tau^k$ moves the three nodes around a single loop. Starting from an input state $u$ of node 1, three applications of $g$ visit every node once and return to node 1 at the input state $u\oplus b_1\oplus b_2\oplus b_3$, since every node is the input of exactly one node. During these three applications, the output is complemented $|b|=b_1+b_2+b_3$ times, once for every flipped node. If $|b|$ is even, the six input states form $C(g)=2$ cycles of length three, each with $|b|$ complements, so $|\cM^g|=2^2=4$. If $|b|$ is odd, an input state of node 1 first returns as its complement, and the six input states form a single cycle of length six, with $2|b|$ complements. Thus, $C(g)=1$ and $|\cM^g|=2$. In both cases every cycle is even, because the number of complemented inputs and the number of complemented outputs around the loop are both equal to $|b|$. Since four of the eight flip vectors have even $|b|$, each of the two rotations contributes $4\cdot4+4\cdot2=24$.
\begin{table}[t]
\caption{Contributions to Eq.~\eqref{eq:orbit-count-S} for the directed ring with $N=3$. For each rotation $\tau^k$, the node-flips that leave some logic matrix unchanged are listed with the number $C(g)$ of cycles of input states, and the last column is the sum of $2^{C(g)}$ over these node-flips.}
\label{tab:ring3}
\begin{ruledtabular}
\begin{tabular}{lccc}
$k$ & Node-flips & $C(g)$ & Sum\\
\hline
0 & $b=000$ / $b=111$ & 6/3 & 72\\
1, 2 & $|b|$ even / $|b|$ odd (4/4) & 2/1 & 24\\
\end{tabular}
\end{ruledtabular}
\end{table}
Table~\ref{tab:ring3} summarizes these contributions, and Eq.~\eqref{eq:orbit-count-S} gives
\begin{equation}
|\cE/\GFP|=\frac{64+8+2\cdot24}{2^3\cdot3}=\frac{120}{24}=5.
\label{eq:ring3-orbits}
\end{equation}

The five orbits can also be identified directly. Flipping node $j$ exchanges the constants 0 and 1, or the identity and the negation, at node $j$, and it exchanges the identity and the negation at node $j+1$ if $f_{j+1}$ is non-constant. Thus, the node-flips never change which BFs are non-constant, and the number $n$ of non-constant BFs is invariant under $\GFP$. For $n\leq2$, the node-flips reach every assignment of BFs with the same non-constant nodes, and since any two nodes of the ring are adjacent, the rotations make the positions of the non-constant nodes irrelevant. For $n=3$, every node-flip exchanges the identity and the negation at two nodes, so the parity of the number of negations is also invariant. Therefore, the orbits are those with $n=0$, $1$, and $2$, and those with $n=3$ and an even or an odd number of negations. They contain 16, 48, 48, 8, and 8 of the 128 representations, respectively. Direct enumeration of the expanded ensemble gives the same five orbits.

The SSNs of the five orbits are pairwise non-isomorphic. For $n\leq2$, every state reaches a single fixed point, and the longest transient has $n+1$ steps. For $n=3$, the BN is a permutation of the states. With an even number of negations, it rotates the state, so that $\ket{000}$ and $\ket{111}$ are fixed points and the other six states form two cycles of length three. With an odd number of negations, the states form one cycle of length six and one of length two. Thus, for the ring, the orbits of $\GFP$ are exactly the classes of BNs with isomorphic SSNs. This is consistent with, though not implied by, Theorem~\ref{thm:GFP-minimal}, according to which no state permutation outside $\GFP$ is admissible for every BN on the ring.

As a second check, for the fully connected structure without self-inputs, in which every node reads every other node, the orbits of $\GFP$ are the structural equivalence classes of Ref.~\cite{glass1975classification}, since without self-inputs the asynchronous diagram of Ref.~\cite{glass1975classification} determines the BN, and a hypercube symmetry maps it rigidly onto the diagram of the transformed BN~\cite{fabre-monplaisir_isometries_2021}. Direct enumeration gives 4 orbits for $N = 2$ and 112 orbits of the 4096 BNs for $N = 3$, in agreement with Glass’s counts.

\bibliographystyle{apsrev4-2}
\FloatBarrier
\
\bibliography{references}

@book{Boole1847,
  author    = {Boole, George},
  title     = {The Mathematical Analysis of Logic: Being an Essay Towards a Calculus of Deductive Reasoning},
  publisher = {Macmillan, Barclay, and Macmillan},
  address   = {Cambridge},
  year      = {1847}
}

@book{Schroeder1890,
  author    = {Schr{\"o}der, Ernst},
  title     = {Vorlesungen {\"u}ber die Algebra der Logik (Exakte Logik)},
  volume    = {1--3},
  publisher = {B. G. Teubner},
  address   = {Leipzig},
  year      = {1890--1905}
}

@book{Couturat1905,
  author    = {Couturat, Louis},
  title     = {L'Alg{\`e}bre de la logique},
  series    = {Scientia},
  number    = {24},
  publisher = {Gauthier-Villars},
  address   = {Paris},
  year      = {1905},
  note      = {{Russian} translation by I. Sle{\v s}inskij (Mathesis, Odessa, 1909); {English} translation by L. G. Robinson, \textit{The Algebra of Logic} (Open Court, Chicago, 1914).}
}

@article{Minkowski_1909,
  author  = {Minkowski, Hermann},
  title   = {Raum und {Zeit}},
  journal = {Phys. Z.},
  volume  = {10},
  pages   = {104--111},
  year    = {1909}
}

@article{Ehrenfest1910,
  author  = {Ehrenfest, Paul},
  title   = {Review of {L. Couturat}, \textit{{The Algebra of Logic}} ({Russian} translation, {Odessa}, 1909)},
  journal = {Zh. Russ. Fiz.-Khim. O-va., Fiz. Otd.},
  volume  = {42},
  pages   = {382--387},
  year    = {1910},
  note    = {in {Russian}; {French} translation: P. Ehrenfest, Compte rendu de: Louis Couturat, \textit{L'Alg{\`e}bre de la logique}, Philosophia Scienti{\ae} \textbf{15}(2), 179 (2011).}
}

@article{Nakashima1936,
  author  = {Nakashima, Akira},
  title   = {The theory of relay circuit composition},
  journal = {Nippon Electr. Commun. Eng.},
  volume  = {3},
  pages   = {197--226},
  year    = {1936},
  note    = {translated from the {Japanese} original in J. Inst. Telegr. Teleph. Eng. Jpn. \textbf{150}, 731--752 (1935)}
}

@article{Polya1937,
  author    = {P{\'o}lya, George},
  title     = {Kombinatorische Anzahlbestimmungen f{\"u}r Gruppen, Graphen und chemische Verbindungen},
  journal   = {Acta Mathematica},
  volume    = {68},
  number    = {1},
  pages     = {145--254},
  year      = {1937},
  doi       = {10.1007/BF02546665},
}

@article{shannon_symbolic_1938,
  author  = {Shannon, Claude E.},
  title   = {A symbolic analysis of relay and switching circuits},
  journal = {Trans. Am. Inst. Electr. Eng.},
  volume  = {57},
  number  = {12},
  pages   = {713--723},
  year    = {1938},
  doi     = {10.1109/T-AIEE.1938.5057767}
}

@article{Shestakov1941,
  author  = {Shestakov, Victor I.},
  title   = {Algebra of two-pole schemes constructed exclusively from two-pole elements (algebra of {A}-schemes)},
  journal = {Zh. Tekh. Fiz.},
  volume  = {11},
  number  = {6},
  pages   = {532--549},
  year    = {1941},
  note    = {in {Russian}; a shortened version appeared in Avtomat. Telemekh. No.~2, 15--24 (1941)}
}

@article{Waddington1942,
  author  = {Waddington, C. H.},
  title   = {Canalization of development and the inheritance of acquired characters},
  journal = {Nature},
  volume  = {150},
  pages   = {563--565},
  year    = {1942},
  doi     = {10.1038/150563a0}
}

@article{turing1952chemical,
  author  = {Turing, Alan M.},
  title   = {The chemical basis of morphogenesis},
  journal = {Philos. Trans. R. Soc. London B},
  volume  = {237},
  number  = {641},
  pages   = {37--72},
  year    = {1952},
  doi     = {10.1098/rstb.1952.0012}
}

@article{Harrison1963,
  author  = {Harrison, Michael A.},
  title   = {The number of transitivity sets of {Boolean} functions},
  journal = {J. Soc. Ind. Appl. Math.},
  volume  = {11},
  number  = {3},
  pages   = {806--828},
  year    = {1963},
  doi     = {10.1137/0111059}
}

@book{Harrison1965,
  author    = {Harrison, Michael A.},
  title     = {Introduction to Switching and Automata Theory},
  publisher = {McGraw-Hill},
  address   = {New York},
  year      = {1965}
}

@article{Kauffman1969,
  author  = {Kauffman, Stuart A.},
  title   = {Metabolic stability and epigenesis in randomly constructed genetic nets},
  journal = {J. Theor. Biol.},
  volume  = {22},
  number  = {3},
  pages   = {437--467},
  year    = {1969},
  doi     = {10.1016/0022-5193(69)90015-0}
}

@misc{Turing1969,
  author       = {Turing, Alan M.},
  title        = {{Intelligent Machinery}},
  howpublished = {in \textit{Machine Intelligence 5}, edited by B. Meltzer and D. Michie (Edinburgh University Press, Edinburgh, 1969) pp.~3--23},
  note         = {report written for the {National Physical Laboratory} in 1948}
}

@article{glass1975classification,
  author  = {Glass, Leon},
  title   = {Classification of biological networks by their qualitative dynamics},
  journal = {J. Theor. Biol.},
  volume  = {54},
  number  = {1},
  pages   = {85--107},
  year    = {1975},
  doi     = {10.1016/S0022-5193(75)80056-7}
}

@book{robert_discrete_1986,
  author    = {Robert, Fran{\c{c}}ois},
  title     = {Discrete Iterations: A Metric Study},
  series    = {Springer Series in Computational Mathematics},
  volume    = {6},
  publisher = {Springer},
  address   = {Berlin, Heidelberg},
  year      = {1986},
  doi       = {10.1007/978-3-642-61607-5}
}

@book{Brown1990,
  author    = {Brown, Frank Markham},
  title     = {Boolean Reasoning: The Logic of {Boolean} Equations},
  publisher = {Kluwer Academic Publishers},
  address   = {Boston},
  year      = {1990}
}

@article{burch_symbolic_1992,
  author  = {Burch, Jerry R. and Clarke, Edmund M. and McMillan, Kenneth L. and Dill, David L. and Hwang, L. J.},
  title   = {Symbolic model checking: {$10^{20}$} states and beyond},
  journal = {Inf. Comput.},
  volume  = {98},
  number  = {2},
  pages   = {142--170},
  year    = {1992},
  doi     = {10.1016/0890-5401(92)90017-A}
}

@book{Kauffman1993,
  author    = {Kauffman, Stuart A.},
  title     = {The Origins of Order: Self-Organization and Selection in Evolution},
  publisher = {Oxford University Press},
  address   = {New York},
  year      = {1993}
}

@article{Maini1997SpatialPatterns,
  author  = {Maini, Philip K. and Painter, Kevin J. and Chau, Helene Nguyen Phong},
  title   = {Spatial pattern formation in chemical and biological systems},
  journal = {J. Chem. Soc., Faraday Trans.},
  volume  = {93},
  number  = {20},
  pages   = {3601--3610},
  year    = {1997},
  doi     = {10.1039/A702602A}
}

@article{Bornholdt2000,
  author  = {Bornholdt, Stefan and Rohlf, Thimo},
  title   = {Topological evolution of dynamical networks: Global criticality from local dynamics},
  journal = {Phys. Rev. Lett.},
  volume  = {84},
  number  = {26},
  pages   = {6114--6117},
  year    = {2000},
  doi     = {10.1103/PhysRevLett.84.6114}
}

@article{laubenbacher_equivalence_2001,
  author  = {Laubenbacher, Reinhard and Pareigis, Bodo},
  title   = {Equivalence relations on finite dynamical systems},
  journal = {Adv. Appl. Math.},
  volume  = {26},
  number  = {3},
  pages   = {237--251},
  year    = {2001},
  doi     = {10.1006/aama.2000.0717}
}

@article{bassler_evolution_2004,
  author  = {Bassler, Kevin E. and Lee, Choongseop and Lee, Yong},
  title   = {Evolution of developmental canalization in networks of competing {Boolean} nodes},
  journal = {Phys. Rev. Lett.},
  volume  = {93},
  number  = {3},
  pages   = {038101},
  year    = {2004},
  doi     = {10.1103/PhysRevLett.93.038101}
}

@article{li_yeast_2004,
  author  = {Li, Fangting and Long, Tao and Lu, Ying and Ouyang, Qi and Tang, Chao},
  title   = {The yeast cell-cycle network is robustly designed},
  journal = {Proc. Natl. Acad. Sci. U.S.A.},
  volume  = {101},
  number  = {14},
  pages   = {4781--4786},
  year    = {2004},
  doi     = {10.1073/pnas.0305937101}
}

@article{golubitsky_nonlinear_2006,
  author  = {Golubitsky, Martin and Stewart, Ian},
  title   = {Nonlinear dynamics of networks: The groupoid formalism},
  journal = {Bull. Am. Math. Soc.},
  volume  = {43},
  number  = {3},
  pages   = {305--365},
  year    = {2006},
  doi     = {10.1090/S0273-0979-06-01108-6}
}

@article{liu_emergent_2006,
  author  = {Liu, Min and Bassler, Kevin E.},
  title   = {Emergent criticality from coevolution in random {Boolean} networks},
  journal = {Phys. Rev. E},
  volume  = {74},
  number  = {4},
  pages   = {041910},
  year    = {2006},
  doi     = {10.1103/PhysRevE.74.041910}
}

@article{Sorrentino_2007,
  author  = {Sorrentino, Francesco and Ott, Edward},
  title   = {Network synchronization of groups},
  journal = {Phys. Rev. E},
  volume  = {76},
  number  = {5},
  pages   = {056114},
  year    = {2007},
  doi     = {10.1103/PhysRevE.76.056114}
}

@article{reichhardt_canalization_2007,
  author  = {Reichhardt, Cynthia J. Olson and Bassler, Kevin E.},
  title   = {Canalization and symmetry in {Boolean} models for genetic regulatory networks},
  journal = {J. Phys. A: Math. Theor.},
  volume  = {40},
  number  = {16},
  pages   = {4339--4350},
  year    = {2007},
  doi     = {10.1088/1751-8113/40/16/006}
}

@article{davidich_boolean_2008,
  author  = {Davidich, Maria I. and Bornholdt, Stefan},
  title   = {{Boolean} network model predicts cell cycle sequence of fission yeast},
  journal = {PLoS ONE},
  volume  = {3},
  number  = {2},
  pages   = {e1672},
  year    = {2008},
  doi     = {10.1371/journal.pone.0001672}
}

@article{Garg2008,
  author  = {Garg, Abhishek and Xenarios, Ioannis and Mendoza, Luis and De Micheli, Giovanni},
  title   = {Synchronous versus asynchronous modeling of gene regulatory networks},
  journal = {Bioinformatics},
  volume  = {24},
  number  = {17},
  pages   = {1917--1925},
  year    = {2008},
  doi     = {10.1093/bioinformatics/btn336}
}

@article{Goryachev2008Cdc42,
  author  = {Goryachev, Andrew B. and Pokhilko, Alexandra V.},
  title   = {Dynamics of {Cdc42} network embodies a {Turing}-type mechanism of yeast cell polarity},
  journal = {FEBS Lett.},
  volume  = {582},
  number  = {10},
  pages   = {1437--1443},
  year    = {2008},
  doi     = {10.1016/j.febslet.2008.03.029}
}

@article{Xiao_symmetry_2008,
  author  = {Xiao, Yanghua and Xiong, Momiao and Wang, Wei and Wang, Hui},
  title   = {Emergence of symmetry in complex networks},
  journal = {Phys. Rev. E},
  volume  = {77},
  number  = {6},
  pages   = {066108},
  year    = {2008},
  doi     = {10.1103/PhysRevE.77.066108}
}

@article{irons_logical_2009,
  author  = {Irons, D. J.},
  title   = {Logical analysis of the budding yeast cell cycle},
  journal = {J. Theor. Biol.},
  volume  = {257},
  number  = {4},
  pages   = {543--559},
  year    = {2009},
  doi     = {10.1016/j.jtbi.2008.12.028}
}

@article{szejka_evolution_2010,
  author  = {Szejka, Agnes and Drossel, Barbara},
  title   = {Evolution of {Boolean} networks under selection for a robust response to external inputs yields an extensive neutral space},
  journal = {Phys. Rev. E},
  volume  = {81},
  number  = {2},
  pages   = {021908},
  year    = {2010},
  doi     = {10.1103/PhysRevE.81.021908}
}

@article{liu_finite_2011,
  author  = {Liu, Min and Bassler, Kevin E.},
  title   = {Finite size effects and symmetry breaking in the evolution of networks of competing {Boolean} nodes},
  journal = {J. Phys. A: Math. Theor.},
  volume  = {44},
  number  = {4},
  pages   = {045101},
  year    = {2011},
  doi     = {10.1088/1751-8113/44/4/045101}
}

@article{matthew_d_reichl_canalization_2011,
  author  = {Reichl, Matthew D. and Bassler, Kevin E.},
  title   = {Canalization in the critical states of highly connected networks of competing {Boolean} nodes},
  journal = {Phys. Rev. E},
  volume  = {84},
  number  = {5},
  pages   = {056103},
  year    = {2011},
  doi     = {10.1103/PhysRevE.84.056103}
}

@book{cheng2011analysis,
  author    = {Cheng, Daizhan and Qi, Hongsheng and Li, Zhiqiang},
  title     = {Analysis and Control of {Boolean} Networks: A Semi-Tensor Product Approach},
  publisher = {Springer},
  address   = {London},
  year      = {2011},
  doi       = {10.1007/978-0-85729-097-7}
}

@article{MarquesPita2013,
  author  = {Marques-Pita, Manuel and Rocha, Luis M.},
  title   = {Canalization and control in automata networks: Body segmentation in \textit{{Drosophila melanogaster}}},
  journal = {PLoS ONE},
  volume  = {8},
  number  = {3},
  pages   = {e55946},
  year    = {2013},
  doi     = {10.1371/journal.pone.0055946}
}

@article{snyder_computational_2013,
  author  = {Snyder, David and Goudarzi, Alireza and Teuscher, Christof},
  title   = {Computational capabilities of random automata networks for reservoir computing},
  journal = {Phys. Rev. E},
  volume  = {87},
  number  = {4},
  pages   = {042808},
  year    = {2013},
  doi     = {10.1103/PhysRevE.87.042808}
}

@article{shabnam_hossein_symmetry_2014,
  author  = {Hossein, Shabnam and Reichl, Matthew D. and Bassler, Kevin E.},
  title   = {Symmetry in critical random {Boolean} network dynamics},
  journal = {Phys. Rev. E},
  volume  = {89},
  number  = {4},
  pages   = {042808},
  year    = {2014},
  doi     = {10.1103/PhysRevE.89.042808}
}

@article{nyberg2015mesoscopic,
  author  = {Nyberg, Amy and Gross, Thilo and Bassler, Kevin E.},
  title   = {Mesoscopic structures and the {Laplacian} spectra of random geometric graphs},
  journal = {J. Complex Netw.},
  volume  = {3},
  number  = {4},
  pages   = {543--551},
  year    = {2015},
  doi     = {10.1093/comnet/cnv004}
}

@article{abou_jaoude_logical_2016,
  author  = {Abou-Jaoud{\'e}, Wassim and Traynard, Pauline and Monteiro, Pedro T. and Saez-Rodriguez, Julio and Helikar, Tom{\'a}{\v s} and Thieffry, Denis and Chaouiya, Claudine},
  title   = {Logical modeling and dynamical analysis of cellular networks},
  journal = {Front. Genet.},
  volume  = {7},
  pages   = {94},
  year    = {2016},
  doi     = {10.3389/fgene.2016.00094}
}

@article{gadouleau_influence_2020,
  author  = {Gadouleau, Maximilien},
  title   = {On the influence of the interaction graph on a finite dynamical system},
  journal = {Nat. Comput.},
  volume  = {19},
  number  = {1},
  pages   = {15--28},
  year    = {2020},
  doi     = {10.1007/s11047-019-09732-y}
}

@article{SanchezGarcia_Symmetry_2020,
  author  = {S{\'a}nchez-Garc{\'\i}a, Rub{\'e}n J.},
  title   = {Exploiting symmetry in network analysis},
  journal = {Commun. Phys.},
  volume  = {3},
  number  = {1},
  pages   = {82},
  year    = {2020},
  doi     = {10.1038/s42005-020-0345-z}
}

@article{fabre-monplaisir_isometries_2021,
  author  = {Fabre-Monplaisir, Jean and Moss{\'e}, Brigitte and Remy, {\'E}lisabeth},
  title   = {Isometries of the hypercube: A tool for {Boolean} regulatory networks analysis},
  journal = {Physica D},
  volume  = {424},
  pages   = {132831},
  year    = {2021},
  doi     = {10.1016/j.physd.2020.132831}
}

@article{aledo_symmetrizable_2023,
  author  = {Aledo, Juan A. and Goles, Eric and Montalva-Medel, Marco and Montealegre, Pedro and Valverde, Jos{\'e} C.},
  title   = {Symmetrizable {Boolean} networks},
  journal = {Inf. Sci.},
  volume  = {626},
  pages   = {787--804},
  year    = {2023},
  doi     = {10.1016/j.ins.2023.01.082}
}

@article{bridoux_interaction_2023,
  author  = {Bridoux, Florian and Perrot, K{\'e}vin and Picard Marchetto, Aymeric and Richard, Adrien},
  title   = {Interaction graphs of isomorphic automata networks {I}: Complete digraph and minimum in-degree},
  journal = {J. Comput. Syst. Sci.},
  volume  = {138},
  pages   = {103458},
  year    = {2023},
  doi     = {10.1016/j.jcss.2023.05.003}
}

@article{kadelka_meta-analysis_2024,
  author  = {Kadelka, Claus and Butrie, Taras-Michael and Hilton, Evan and Kinseth, Jack and Schmidt, Addison and Serdarevic, Haris},
  title   = {A meta-analysis of {Boolean} network models reveals design principles of gene regulatory networks},
  journal = {Sci. Adv.},
  volume  = {10},
  number  = {2},
  pages   = {eadj0822},
  year    = {2024},
  doi     = {10.1126/sciadv.adj0822}
}

@book{tinkham_group_1964,
  author    = {Tinkham, Michael},
  title     = {Group Theory and Quantum Mechanics},
  publisher = {McGraw-Hill},
  address   = {New York},
  year      = {1964}
}

@article{noether_invariant_1918,
  author  = {Noether, Emmy},
  title   = {Invariante {Variationsprobleme}},
  journal = {Nachr. Ges. Wiss. G{\"o}ttingen, Math.-Phys. Kl.},
  volume  = {1918},
  pages   = {235--257},
  year    = {1918},
  url     = {https://eudml.org/doc/59024}
}

@book{marsden_ratiu_1999,
  author    = {Marsden, Jerrold E. and Ratiu, Tudor S.},
  title     = {Introduction to Mechanics and Symmetry: A Basic Exposition of Classical Mechanical Systems},
  edition   = {2},
  series    = {Texts in Applied Mathematics},
  volume    = {17},
  publisher = {Springer},
  address   = {New York},
  year      = {1999},
  doi       = {10.1007/978-0-387-21792-5}
}

@article{landau_phase_1937,
  author  = {Landau, L. D.},
  title   = {On the theory of phase transitions},
  journal = {Zh. Eksp. Teor. Fiz.},
  volume  = {7},
  pages   = {19--32},
  year    = {1937},
  note    = {{English} translation in \textit{Collected Papers of L. D. Landau} (Pergamon, Oxford, 1965), pp.~193--216},
  doi     = {10.1016/B978-0-08-010586-4.50034-1}
}

@article{wilson_kogut_1974,
  author  = {Wilson, Kenneth G. and Kogut, J.},
  title   = {The renormalization group and the {$\epsilon$} expansion},
  journal = {Phys. Rep.},
  volume  = {12},
  number  = {2},
  pages   = {75--199},
  year    = {1974},
  doi     = {10.1016/0370-1573(74)90023-4}
}

@article{yang_mills_1954,
  author  = {Yang, C. N. and Mills, R. L.},
  title   = {Conservation of isotopic spin and isotopic gauge invariance},
  journal = {Phys. Rev.},
  volume  = {96},
  number  = {1},
  pages   = {191--195},
  year    = {1954},
  doi     = {10.1103/PhysRev.96.191}
}

@article{weinberg_leptons_1967,
  author  = {Weinberg, Steven},
  title   = {A model of leptons},
  journal = {Phys. Rev. Lett.},
  volume  = {19},
  number  = {21},
  pages   = {1264--1266},
  year    = {1967},
  doi     = {10.1103/PhysRevLett.19.1264}
}

@book{golubitsky_stewart_schaeffer_1988,
  author    = {Golubitsky, Martin and Stewart, Ian and Schaeffer, David G.},
  title     = {Singularities and Groups in Bifurcation Theory, Volume {II}},
  series    = {Applied Mathematical Sciences},
  volume    = {69},
  publisher = {Springer},
  address   = {New York},
  year      = {1988},
  doi       = {10.1007/978-1-4612-4574-2}
}

@article{stewart_groupoids_2003,
  author  = {Stewart, Ian and Golubitsky, Martin and Pivato, Marcus},
  title   = {Symmetry groupoids and patterns of synchrony in coupled cell networks},
  journal = {SIAM J. Appl. Dyn. Syst.},
  volume  = {2},
  number  = {4},
  pages   = {609--646},
  year    = {2003},
  doi     = {10.1137/S1111111103419896}
}

@article{pecora_cluster_2014,
  author  = {Pecora, Louis M. and Sorrentino, Francesco and Hagerstrom, Aaron M. and Murphy, Thomas E. and Roy, Rajarshi},
  title   = {Cluster synchronization and isolated desynchronization in complex networks with symmetries},
  journal = {Nat. Commun.},
  volume  = {5},
  pages   = {4079},
  year    = {2014},
  doi     = {10.1038/ncomms5079}
}

@article{toner_flocks_1998,
  author  = {Toner, John and Tu, Yuhai},
  title   = {Flocks, herds, and schools: A quantitative theory of flocking},
  journal = {Phys. Rev. E},
  volume  = {58},
  number  = {4},
  pages   = {4828--4858},
  year    = {1998},
  doi     = {10.1103/PhysRevE.58.4828}
}

@article{solon_flocking_2015,
  author  = {Solon, A. P. and Tailleur, J.},
  title   = {Flocking with discrete symmetry: The two-dimensional active {Ising} model},
  journal = {Phys. Rev. E},
  volume  = {92},
  number  = {4},
  pages   = {042119},
  year    = {2015},
  doi     = {10.1103/PhysRevE.92.042119}
}

\end{document}